\documentclass[a4paper,12pt]{article}
\usepackage{jcappub}
\usepackage{graphicx}
\usepackage{aas_macros}
\usepackage{ulem}
\input{colordvi.tex}
\title{
{
Evolution of perturbations and cosmological constraints in decaying dark matter models with arbitrary decay mass products}}

\author[1]{Shohei Aoyama,}
\author[1,2]{Toyokazu Sekiguchi,}
\author[3]{Kiyotomo Ichiki}
\author[1,3,4]{and Naoshi Sugiyama}

\affiliation[1]{Department of Physics and Astrophysics, Nagoya University, 
Nagoya 464-8602, Japan}
\affiliation[2]{University of Helsinki and Helsinki Institute of Physics, 
P.O. Box 64, FI-00014, Helsinki, Finland}
\affiliation[3]{Kobayashi-Maskawa Institute for the Origin of Particles 
and the Universe, Nagoya University, Nagoya 464-8602, Japan}
\affiliation[4]{Kavli Institute for the Physics and Mathematics of the Universe 
(Kavli IPMU), The University of Tokyo, Chiba 277-8582, Japan}

\abstract{Decaying dark matter (DDM) is a candidate which can
solve the discrepancies between predictions of the concordance
$\Lambda$CDM model and observations at small scales such as the number
counts of companion galaxies of the Milky Way and the density profile at
the center of galaxies.  
Previous studies are limited to the cases where the decay particles are
massless and/or have almost degenerate masses with that of mother
particles. Here we expand the DDM models so that one can consider the
DDM with arbitrary lifetime and the decay products with arbitrary
masses.  We calculate the
time evolutions of perturbed phase-space distribution functions of decay
products for the first time and study effects of DDM on the temperature
anisotropy in the cosmic microwave background and the matter power
spectrum at present.  From a recent observational estimate of
$\sigma_{8}$, we derive constraints on the lifetime of DDM and the mass
ratio between the decay products and DDM.  We also discuss implications
of the DDM model for the discrepancy in the measurements of $\sigma_8$ 
recently claimed by {the Planck satellite collaboration}.}


\begin{document}
\maketitle

\section{Introduction}
Numerous kinds of astrophysical observations suggest the existence of
dark matter, which has no electro-magnetic or strong interactions with
other particles in the Universe.  It is necessary for explaining the
observed phenomena such as flat rotation curves of galaxies,
anisotropies in the cosmic microwave background (CMB) and the formation
of large scale structure within the age of the Universe, 
$t_{\rm age}=13.8$\,Gyr \cite{2013arXiv1303.5062P}{.} 
In particular, to explain
the hierarchical formation of the cosmic structure, dark matter should
have very small momentum {hence} so-called cold dark
matter (CDM) {becomes}
the standard model of dark matter.

However, while it is
widely believed that dark matter consists of particles beyond the
Standard Model, we have little knowledge on the nature of dark matter.
Thus, for example, 
it remains possible that dark matter decays into
other particles with smaller masses.  This kind of dark matter is called
decaying dark matter (DDM).  By constraining the lifetime of DDM, we can
obtain the nature of dark matter and information about a particle theory
lying behind it. {While the CDM model is widely consistent 
with observations on cosmological scales, 
there exist some discrepancies at sub-galactic scales 
between the model predictions by N-body simulations 
and astronomical observations. 
Well known examples are the "cusp problem" {\cite{2004MNRAS.351..903G,2009A&A...505....1V}}, 
i.e., the over-concentration problem of the inner core of galaxies, and the "missing satellite problem" \cite{1999ApJ...524L..19M} or the "too big to fail problem" \cite{2012JCAP...12..007P}, where the number or properties of satellite galaxies of the Milky Way do not match the results from numerical simulations. 
DDM is often discussed as a remedy to such small-scale CDM problems, for example, see Cen \cite{2001ApJ...546L..77C}. The DDM model, however, should pass the tests from precise cosmological observations, such as measurements of the CMB and the large scale structure of the universe. One of the aims of this paper is therefore to give a complete set of perturbation equations of the DDM model with decay products having arbitrary masses, and show some evolutions of density perturbations in the DDM model to compare them with recent cosmological observations.}

Cosmological implications of DDM 
have been studied by several authors
\cite{2001ApJ...546L..77C,2004PhRvL..93g1302I,2003ApJ...597..645O,
2010PhRvD..81j3501P,2005PhRvD..72f3510K,1999PhRvD..60l3508K,
1993NuPhB.403..671K,2009JCAP...06..005D,2008PhRvD..77b3009A,
2005PhR...405..279B,2005PhLB..625....7K,2010PhRvD..81h3511P,2011PhRvD..83f3504B,
2011PhLB..701..530H,2012JCAP...10..017E,2012PhRvD..85d3514W}.
Among them, Kaplinghat \cite{2005PhRvD..72f3510K} and Huo
\cite{2011PhLB..701..530H} have considered models in which dark matter
decays into two "daughter" particles\footnote{We call progenitor
particle "mother" particle.} well before the matter-radiation equality epoch.
They calculated time evolution of perturbations and 
set constraints on 
the mass ratio and 
the life time of DDM from the free streaming scale 
of daughter particles.
The former {considered the DDM which decays into a massive 
particle and massless one in the early Universe. They }
computed the matter power spectrum and discussed the
effects of free-streaming of the decay products 
{by considering the phase-space density of the massive daughter particle}. 
The latter computed the CMB power spectrum
in addition to the matter power spectrum. Moreover, DDM models in which 
the mass difference between the mother and the massive daughter particles 
$\Delta m$ is much smaller than the mass of the mother particles are studied in
refs.~\cite{2010PhRvD..81h3511P,2011PhRvD..83f3504B,2012PhRvD..85d3514W,2013PhRvD..88l3515W}.
In particular, Peter \cite{2010PhRvD..81h3511P} has
considered effects of the recoil energy of the daughter particles on 
the halo mass-concentration and galaxy-cluster mass function 
using N-body simulations and set 
constraints on the lifetime of DDM and $\Delta m$
in terms of the kick velocity {$v_{\rm kick}$} by demanding that the daughter particles in
halos with mass $M=10^{12}M_{\odot}$ do not escape from the halos and
destroy the gravitational potential.
{She} also set constraints 
by using weak lensing measurements and X-ray observations.
{Wang et al. have analyzed N-body simulations with 
DDM and focused on the statistical properties of 
the transmitted Lyman-$\alpha$ (Ly-$\alpha$) forest flux. 
By comparing the data of the Sloan Digital Sky Survey (SDSS) 
and WMAP 7 with their simulation data, 
they set a constraint on the life time of DDM {as}
$\Gamma^{-1}>40$ Gyr where $v_{\rm kick}\gtrsim 50$ km/s \cite{2013PhRvD..88l3515W}.}
{However, previous studies so far could not deal with DDM models
where mother particles decay into particles with arbitrary mass at late times (i.e. after the matter-radiation equality).
This is what we will explore in this paper.}

In this paper, 
{we investigate models of DDM which decays into two daughter particles.}
We study linear cosmological perturbations in the DDM model without
approximations such that the decay products are massless or highly
non-relativistic. Instead, we
directly solve the Boltzmann equations for the mother and daughter
particles and follow the time evolutions of their distribution functions in
the discretized phase space. In this formulation,
one can set masses of both daughter particles to arbitrary values.
{While our formulation and numerical implementation
can deal with arbitrary lifetime and/or masses of daughter particles, 
to minimize complication, we assume that DDM decays after cosmological recombination 
into two particles one of which has {a} finite mass and the other is massless.}

Here, we basically follow the notation of Ma and Bertschinger
\cite{1995ApJ...455....7M}. We adopt natural units with $c=\hbar=k_{\rm
B}=1$. The proper and conformal time are denoted by $t$ and $\tau$, respectively.
A dot represents partial derivative with respect to the conformal time,
i.e., $\dot f\equiv  \partial f/\partial \tau$. A quantity with a subscript
"$\ast$" indicates its value at cosmological recombination.

The layout of this paper is as follows. 
{In the following section, 
we introduce our DDM model and present the Boltzmann equations, 
which describe the perturbation evolutions. 
In addition, we also describe how
we solve the equations numerically.  
Section \ref{sec:evol} denotes 
time evolutions of perturbations in the DDM model. 
Effects on cosmological observables including 
the angular power spectrum of the CMB anisotropy $C_l$ 
and the matter power spectrum
$P(k)$ are discussed in section \ref{sec:signatures}. 
In section \ref{sec:discussion}, 
we discuss some interpretations for the effects of
the decay of dark matter and derive a
constraint on the DDM model from an 
observed value of $\sigma_{8}$. 
Finally, section \ref{sec:conclusion} conclude{s} this paper.
}

\section{Model of DDM and Boltzmann equation}\label{sec:model}
\subsection{DDM model and set up}\label{all.2}

In this work, we consider a model of DDM ({\sf M}) 
which decays into two particles ({\sf D1}, {\sf D2})
well after cosmological recombination, 
\begin{equation} 
{\sf M}\to {\sf D1} +{\sf D2}~~~({\rm with}~\Gamma^{-1}\gg t_{\ast})~,
\end{equation}
where {$\Gamma$ is the decay rate of DDM.}

{Throughout this paper, we assume a variant of the concordance flat power-law $\Lambda$CDM model
where CDM is replaced with DDM,  and consider cosmological perturbations.
The model is specified with the following cosmological parameters:
\begin{equation}
(\Omega_b,~\Omega_{\rm M\emptyset}, h_\emptyset, \tau_{\rm reion}, n_s, A_s, 
\Gamma, m_{\rm D1}/m_{\rm M}, m_{\rm D2}/m_{\rm M}), 
\label{eq:paramet}
\end{equation}
where $\Omega_b$ and $\Omega_{\rm M\emptyset}$ are the density
parameters of baryons and mother particles, respectively, $h_\emptyset$ is the reduced
Hubble constant estimated assuming that the DDM does not decay, 
$\tau_{\rm reion}$ is the 
optical depth of reionization, and $n_s$ and $A_s$ are respectively the spectral index and 
amplitude of the primordial curvature perturbation at $k =
0.002~$Mpc$^{-1}$.
 Here and in the following, the subscript 
$\emptyset$ indicates quantities which are estimated assuming $\Gamma=0$.
Following the WMAP 7-year results \cite{2011ApJS..192...18K}\footnote{
Recently, the cosmological parameters derived  
by Planck have been reported \cite{2013arXiv1303.5076P}. 
Since the estimated values of $\Omega_{\rm b}h^{2}$ and $\Omega_{\rm c}h^{2}$ change
from those of the WMAP 7-year results only by several percent, 
results and constraints presented in this paper may not be 
affected significantly when the cosmological parameters from Planck are adopted.}, 
we fix $(\Omega_b,~\Omega_{\rm M\emptyset}, h_\emptyset, \tau_{\rm reion}, n_s, A_s)$
to $(0.0454,~0.226,~0.704,~0.088,~0.967, 2.43\times 10^{-9})$. 
With the parameterization eq.~\eqref{eq:paramet}, the genuine Hubble
constant $h$  and the density parameter of mother particles $\Omega_{\rm M}$
are derived parameters which can be obtained by solving the background evolution.
It should be noted that these values of parameters are not necessary 
to be {the} best fitting values {to WMAP-7 year data 
in the DDM model}. These values are adopted only for a reference.
Regarding $m_{\rm D2}/m_{\rm M}$, while the formalism we present in this section 
is applicable for arbitrary masses of the daughter particles, 
{as is stated in introduction,} we in the rest of this paper 
assume that the mass of one of the daughter particles 
can be nonzero and the other one is massless. Thus, 
$\Gamma$ and $m_{\rm D1}/m_{\rm M}$ are treated as free parameters with 
$m_{\rm {D2}}/m_{\rm M}$ being fixed to zero.}
Subscripts "M", "D1" and "D2" indicate 
the mother, massive daughter, and massless daughter particles, respectively.

\subsection{Boltzmann equations}
In this paper, we choose the synchronous gauge of the mother particles
to describe {cosmological linear} perturbations. 
The line element is given by 
\begin{equation} 
ds^{2}=a(\tau)^{2}\left\{d\tau^{2} +
 \left(
  \delta_{ij}+h_{ij}(\boldsymbol x, \tau) 
 \right)
dx^{i}dx^{j}
\right\}~, 
\end{equation}
{where $a$ is the scale factor and $h_{ij}$ represents the metric perturbations.
In terms of Fourier components,  $h_{ij}$ can be given as }
\begin{equation} 
h_{ij}(\boldsymbol x, \tau) =\displaystyle\int 
d^{3}k
\left( h_{\rm L}(\boldsymbol{k},\tau) 
\hat{\boldsymbol{k}}_{i} \hat{\boldsymbol{k}}_{j}
+6\eta_{\rm T} (\boldsymbol{k},\tau) 
(\hat{\boldsymbol{k}}_{i} \hat{\boldsymbol{k}}_{j} - \dfrac{1}{3}\delta_{ij})
\right)
\exp ( i \boldsymbol{k}\cdot \boldsymbol{x})
\label{fourierDefinition}~,
\end{equation}
where $\boldsymbol{k} \equiv k\hat{\boldsymbol{k}}$ 
is a wave number vector and $\hat{\boldsymbol{k}}$ is the unit vector of 
$\boldsymbol{k}$. 
{In the rest of this section, we focus on a single mode of perturbations with $\boldsymbol k$
and often abbreviate dependences of perturbed variables on $\boldsymbol k$.}

The Boltzmann equation which describes time evolution of a phase-space distribution function 
$f(\boldsymbol x, \boldsymbol q, \tau)$ can be written as
\begin{equation} 
 \dfrac{\partial f}{\partial \tau}
+\dfrac{dx^i}{d\tau }\dfrac{\partial f}{\partial x^i}
+\dfrac{dq}{d\tau }\dfrac{\partial f}{\partial q}
+\dfrac{dn^i}{d\tau }\dfrac{\partial f}{\partial n^i}
=\left(\dfrac{\partial f}{\partial \tau } \right)_{\rm C}~,
\label{BoltzmannGeneral}
\end{equation}
{where $\boldsymbol q$ is the comoving momentum, which is related to the physical momentum $\boldsymbol p$ 
by $\boldsymbol q=a\boldsymbol p$, and $q$ and $n^i=q^i/q$ are respectively the norm and the direction of $\boldsymbol q$. 
For a perturbation mode with $\boldsymbol k$, the second term of the left hand side can be rewritten as 
$i(\hat{\boldsymbol{k}}\cdot \boldsymbol{n})\left(q \slash \varepsilon \right) f$, 
where $\varepsilon $ is  the comoving energy.
In the left hand side, the third term can be rewritten in terms of the metric perturbations by 
adopting the geodesic equation
\begin{equation} 
p^{0}\dfrac{d p^{\mu}}{d \tau}
+{{\Gamma^{\mu}}_{\alpha \beta }}p^{\alpha }p^{\beta }=0~. 
\label{geodesicEquation}
\end{equation}
In particular, time-component ($\mu=0$) of eq.~(\ref{geodesicEquation})
gives  ${d q}\slash{d\tau}$ as
}
\begin{equation} 
\frac{d q}{d\tau}=\dot{\eta}_{\rm T}
-\dfrac{1}{2}\left(\dot{h}_{\rm L}+6\dot{\eta}_{\rm T}
 \right)(\hat{\boldsymbol{k}}\cdot \boldsymbol{n})^{2}
~\label{metricPerturbation}~.
\end{equation}
The fourth term of the left hand side of eq.~(\ref{BoltzmannGeneral}) can be neglected to the first order, because both $\left({dn^{i}}\slash{d\tau}\right)$ and 
$\left({\partial f}\slash{\partial n^{i}}\right)$ are first-order quantities {in the flat Universe}.

The right hand side of eq.~(\ref{BoltzmannGeneral}) is the collision term, 
which {also} describes 
effects of decay or creation of particles on the distribution function. 
Aoyama {et al.}  \cite{2011JCAP...09..025A}
provided  those for the mother  and
daughter particles. 
In this paper, we assume that momenta of the mother particles are negligibly small 
compared with the mass. 
{Under this assumption in conjunction with our choice of gauge, 
the distribution function of the mother particles 
should be proportional to the delta function of $\boldsymbol q$.
Then we obtain}
\begin{equation} 
f_{\rm M}(q, \boldsymbol n, \tau)=N_{\rm M}(\tau)
\delta^{(3)}(\boldsymbol q)~\label{in.other.expression},
\end{equation}
where $N_{\rm M}$ is the {\it comoving} number density of the mother particles. 
Then the collision terms can be recast into \cite{2011JCAP...09..025A}
\begin{eqnarray} 
{\sf M}&:&\left(\frac{\partial f_{\rm M}}{\partial \tau} \right)_{\rm C}=-a\Gamma f_{\rm M}~,
~\label{CollisionTermForMother}\\
{\sf D}m&:&\left(\frac{\partial f_{{\rm D}{m}}}{\partial \tau} \right)_{\rm C}=
\frac{a\Gamma N_{\rm M}}{4\pi q^{2}}
\delta \left(q-ap_{\rm Dmax} \right)
~,\label{CollisionTermForDaughter}
\end{eqnarray}
where $p_{\rm Dmax}$ is the initial physical momentum of decay particles in the rest flame of the mother particles, which
is given by \
\begin{equation} 
p_{\rm Dmax}=\frac12\left[
m_{\rm M}^{2}
-2\left(m_{\rm D1}^{2}+m_{\rm D2}^{2} \right)
+\dfrac{\left(m_{\rm D1}^{2}+m_{\rm D2}^{2} \right)^{2}}{m_{\rm M}^{2}}\right]^{1/2}~.
\end{equation}
Note that collision terms are the same for both the daughter particles.
For later convenience, we introduce $\tau_q$ which is the conformal time when the daughter particles with a comoving momentum $q$
are produced, i.e.,
\begin{equation} 
q=a(\tau_{q})p_{\rm Dmax}~.
\end{equation}

From eqs.~(\ref{metricPerturbation}), (\ref{CollisionTermForMother}) and
(\ref{CollisionTermForDaughter}), the Boltzmann equations for the mother 
and the daughter particles can be written as
\begin{eqnarray} 
{\sf M}&:&\dfrac{\partial f_{\rm M}}{\partial \tau}
+ i\dfrac{qk}{\varepsilon_{\rm M} }(\hat{\boldsymbol{k}}\cdot \boldsymbol{n})f_{\rm M}
+q\dfrac{\partial f_{\rm M}}{\partial q}
\left[\dot{\eta}_{\rm T}-
\dfrac{1}{2}\left(\dot{h}_{\rm L}+6\dot{\eta}_{\rm T} \right)
(\hat{\boldsymbol{k}}\cdot \boldsymbol{n})^{2} \right]=-a\Gamma f_{\rm M}~, 
\label{Boltz.M.1st} \\
{\sf D}m&:&\dfrac{\partial f_{{\rm D}m}}{\partial \tau}
+ i\dfrac{qk}{\varepsilon_{{\rm D}m} }
(\hat{\boldsymbol{k}}\cdot \boldsymbol{n})f_{{\rm D}m}
+q\dfrac{\partial f_{{\rm D}m}}{\partial q}
\left[\dot{\eta}_{\rm T}-
\dfrac{1}{2}\left(\dot{h}_{\rm L}+6\dot{\eta}_{\rm T} \right)
(\hat{\boldsymbol{k}}\cdot \boldsymbol{n})^{2} \right]
\label{Boltz.D.1st} \\
& &=
\dfrac{a \Gamma N_{\rm M} }{4\pi q^{2}}
\delta \left(ap_{\rm Dmax}-q \right) \notag
~,
\end{eqnarray}
where $\varepsilon_{\rm M}$ ($\varepsilon_{{\rm D}m}$)
is the comoving energy of the mother ($m$-th daughter) particles. 
We divide a distribution function $f$
into the background 
$\overline{f}$ 
and the perturbation $\Delta f$ as
\begin{eqnarray} 
f_{\rm M}(q, \boldsymbol{k}, \boldsymbol{n}, \tau)&=&
\overline{f}_{\rm M}(q,\tau)\delta^{(3)}(\boldsymbol{k}) 
+  \Delta f_{\rm M}\left(q,\boldsymbol{k},\boldsymbol{n},\tau \right)~
,\label{perturbM}\\
f_{{\rm D}m}(q, \boldsymbol{k}, \boldsymbol{n}, \tau)&=&
\overline{f}_{{\rm D}m}(q,\tau)\delta^{(3)}(\boldsymbol{k}) 
+\Delta f_{{\rm D}m}\left(q,\boldsymbol{k},\boldsymbol{n},\tau \right)~,\label{perturbD}
\end{eqnarray}
where $\overline{f}$ depends only on $q$ and $\tau $.
Due to the isotropy of the background geometry, $\Delta f$ can be expanded 
in terms of the Legendre polynomials $P_l(\hat{\boldsymbol{k}}\cdot \boldsymbol{n})$ with $l\ge0$.
Therefore we can define $\Delta f_{{\rm M}(l)}$ ($\Delta f_{{\rm D}m(l)}$)
as the $l$-th multipole moment of $\Delta f_{\rm M}$ ($\Delta f_{{\rm D}m}$), i.e. 
\begin{eqnarray} 
\Delta f_{{\rm M}}(q, \boldsymbol{n}, \tau)&=&
\displaystyle\sum_{l=0}^{+\infty}(-i)^{l}(2l+1)\Delta f_{{\rm M}(l)}(q,\tau)
P_{l}(\hat{\boldsymbol{k}}\cdot \boldsymbol{n}),~\label{LegendrePolynomial.M} \\
\Delta f_{{\rm D}m}(q, \boldsymbol{n}, \tau)&=&
\displaystyle\sum_{l=0}^{+\infty}(-i)^{l}(2l+1)\Delta f_{{\rm D}m(l)}(q,\tau)
P_{l}(\hat{\boldsymbol{k}}\cdot \boldsymbol{n})~. \label{LegendrePolynomial.D}
\end{eqnarray}

\subsubsection{Mother particles}

By substituting eq.~(\ref{perturbM}) 
into eq.~(\ref{BoltzmannGeneral}), 
we obtain the Boltzmann equations for the mother particles as follows: 
\begin{eqnarray} 
{\rm unperturbed}&:&\dot{\overline{f}}_{\rm M}=-a\Gamma \overline{f}_{\rm M}~,\label{unperturbedM}\\
{\rm 1st~order }&:&
\dfrac{\partial \Delta f_{\rm M}}{\partial \tau}
+ i\dfrac{qk}{\varepsilon_{\rm M} }(\hat{\boldsymbol{k}}\cdot \boldsymbol{n})\Delta f_{\rm M}
+q\dfrac{\partial \overline{f}_{\rm M}}{\partial q}
\left[\dot{\eta}_{\rm T}-
\dfrac{1}{2}\left(\dot{h}_{\rm L}+6\dot{\eta}_{\rm T} \right)
(\hat{\boldsymbol{k}}\cdot \boldsymbol{n})^{2} \right]\notag \\
& &=-a\Gamma \Delta f_{\rm M}~.
\label{1stPerturbM}
\end{eqnarray}
From eqs.~(\ref{in.other.expression}) and (\ref{unperturbedM}), 
{we obtain 
\begin{equation} 
\overline{f}_{\rm M}(q,t)=\frac{\overline{N}_{\rm M}(\tau)}{4\pi q^2}\delta(q)
\label{fMvar}
\end{equation}
where $\overline N_{\rm M}(\tau)\propto\exp \left(-\Gamma t \right)$ is the mean comoving number density of 
the mother particles. Denoting the mean comoving number density of the mother particles
without decay as $\overline{N}_{{\rm M}\emptyset}$, 
$\overline N_{\rm M}$ can be given as 
\begin{equation}
\overline N_{\rm M}(\tau)=\overline N_{{\rm M}\emptyset}\exp \left(-\Gamma t \right). \label{exponentialCutoff}
\end{equation}
Note that $\overline N_{{\rm M}\emptyset}$ is constant.}

{
According to our gauge choice, the dipole moment of the mother particles is zero. 
Since the mother particles do not have 
nonzero momentum distribution, higher-order multipole moments should vanish, i.e., }
\begin{equation} 
\Delta f_{{\rm M}(l)}=0
~~~({\rm for}~l\ge 1)~.
\label{higherMulti} 
\end{equation}
On the other hand, from eqs.~(\ref{LegendrePolynomial.M}) and (\ref{1stPerturbM}), 
the monopole moment $\Delta f_{\rm M(0)}$ obeys the following equation,
\begin{equation} 
\dfrac{\partial \Delta f_{{\rm M}(0)}}{\partial \tau}
=
\dfrac{1}{6}\dot{h}_{\rm L}
q\dfrac{\partial \overline{f}_{\rm M}}{\partial q}
-a\Gamma \Delta {f}_{\rm M(0)}
~\label{BoltzM}.
\end{equation}

{
Equations \eqref{unperturbedM} and \eqref{BoltzM}
can be recast into evolution equations for the mean energy density $\bar\rho_{\rm M}$
and its perturbation $\bar \rho_{\rm M}\delta_{\rm M}$, which are defined as 
\begin{eqnarray} 
\overline{\rho}_{\rm M}
&\equiv&\frac1{a^4}\displaystyle\int 
dq\,4\pi q^{2} \varepsilon_{\rm M} \overline{f}_{\rm M}~,
\label{meanDensityM} \\
\overline{\rho}_{\rm M}\delta_{\rm M}
&\equiv&\frac1{a^4}\displaystyle\int  
dq\,4\pi q^{2} \varepsilon_{\rm M} \Delta f_{{\rm M}(0)}~.
\label{densityFluctuationM}
\end{eqnarray}
Then by integrating eqs.~\eqref{unperturbedM} and \eqref{BoltzM} multiplied by $4\pi q^2 \varepsilon_{\rm M}/a^4$, 
we obtain
\begin{eqnarray}
\frac{d}{d\tau}\bar\rho_{\rm M}+3\mathcal H \bar\rho_{\rm M}
=-a\Gamma \bar\rho_{\rm M}~, \\
\frac{d}{d\tau}[\bar\rho_{\rm M}\delta_{\rm M}]
+3\mathcal H \bar\rho_{\rm M}\delta_{\rm M}
=-\frac{\dot h_{\rm L}}{2}\bar\rho_{\rm M}
-a\Gamma \bar\rho_{\rm M}\delta_{\rm M}~, 
\end{eqnarray}
where $\mathcal H=\dot a/a$ is the conformal Hubble expansion rate.
In combination, these two equations lead to 
\begin{equation}
\dot{\delta}_{\rm M}=-\frac{\dot h_{\rm L}}2~.\label{deltaM.eq.0.5..h}
\end{equation}
We note that eq.~\eqref{deltaM.eq.0.5..h} is the same as that for CDM 
without decay \cite{1995ApJ...455....7M}. 
We also note that $\bar\rho_{\rm M}$ as $\bar\rho_{\rm M}(\tau)=m_{\rm M}\overline{N}_{\rm M}(\tau)/a^3$
since the mother particles are non-relativistic.
 }

\subsubsection{Daughter particles}

By substituting eq.~(\ref{perturbD}) into eq.~(\ref{Boltz.D.1st}), one can find 
\begin{eqnarray} 
{\rm unperturbed}&:&\dfrac{\partial \overline{f}_{{\rm D}m}}{\partial \tau}
=\dfrac{a\Gamma\overline{N}_{\rm M}}{4\pi q^3\mathcal H} \delta(\tau-\tau_q)
\label{background.D}~, \\
{\rm 1st~order }&:&
\dfrac{\partial \Delta f_{{\rm D}m}}{\partial \tau}
+i\dfrac{qk}{\varepsilon_{{\rm D}m} }
(\hat{\boldsymbol{k}}\cdot \boldsymbol{n})\Delta f_{{\rm D}m}
+q\dfrac{\partial \overline{f}_{{\rm D}m}}{\partial q}
\left[\dot{\eta}_{\rm T}-\dfrac{1}{2}\left(\dot{h}_{\rm L}+6\dot{\eta}_{\rm T} \right)
(\hat{\boldsymbol{k}}\cdot \boldsymbol{n})^{2} \right]\nonumber \\
& &=\dfrac{a\Gamma \overline{N}_{\rm M}}{4\pi q^{3}\mathcal{H}}
\delta_{{\rm M}}
\delta (\tau -\tau_{q})
~.\label{1stPerturbD} 
\end{eqnarray}
In deriving these equations, we used a relation
\begin{equation}
\delta(q-ap_{\rm Dmax})=\frac1{q\mathcal H}\delta(\tau-\tau_q).
\end{equation}
As seen from eq.~\eqref{background.D} and \eqref{1stPerturbD}, 
the unperturbed and first order perturbation equations for both the daughter particles are identical
\cite{2005PhRvD..72f3510K,2012PhRvD..85d3514W,2011JCAP...09..025A}. 
Therefore we can write
\begin{equation} 
\overline{f}_{\rm D1}(q,\tau)=\overline{f}_{\rm D2}(q,\tau)
\equiv \overline{f}_{\rm D}(q,\tau)~.
\end{equation}
As in previous works \cite{2012PhRvD..85d3514W,2011JCAP...09..025A}, 
by solving eq.~\eqref{background.D}, we obtain
\begin{equation} 
\overline{f}_{\rm D}(q,\tau)=
\dfrac{a_{q}\Gamma \overline{N}_{{\rm M}}(\tau_{q})}
{4\pi q^{3} \mathcal{H}_q}
\Theta \left(\tau -\tau_{q} \right)~\label{fD0},
\end{equation}
where $a_{q}=a(\tau_q)$, $\mathcal H_q=\mathcal H(\tau_q)$ 
and $\Theta(x)$ is the Heaviside function. 

Equation (\ref{1stPerturbD}) can be expanded in terms of multipole moments, 
which leads to a Boltzmann hierarchy for the daughter particles: 
\begin{eqnarray} 
\dfrac{\partial }{\partial \tau}(\Delta f_{{\rm D}m(0)})
&=&-\dfrac{q k}{\varepsilon_{{\rm D}m} }
                          \Delta f_{{\rm D}m(1)}
+\dfrac{1}{6}\dot{h}_{\rm L}q
\dfrac{\partial \overline{f}_{\rm D}}{\partial q}
+\dfrac{a\Gamma \overline N_{{\rm M}}}{4\pi q^{3}\mathcal{H}}
\delta_{{\rm M}}
\delta \left(\tau - \tau_{q} \right) ,\label{LegendreD..01} \\
\dfrac{\partial }{\partial \tau}(\Delta f_{{\rm D}m(1)})
&=&\dfrac{q k}{3\varepsilon_{{\rm D}m} }\left( 
 \Delta f_{{\rm D}m(0)}-2\Delta f_{{\rm D}m(2)} \right)~,\label{LegendreD..0101} \\
\dfrac{\partial }{\partial \tau}(\Delta f_{{\rm D}m(2)})~
&=&\dfrac{q k}{5\varepsilon_{{\rm D}m} }\left( 
2\Delta f_{{\rm D}m(1)}-3\Delta f_{{\rm D}m(3)} \right)
-\left(\dfrac{1}{15}\dot{h}_{\rm L}+\dfrac{2}{5}\dot{\eta}_{\rm T} \right)
q\dfrac{\partial \overline{f}_{\rm D}}{\partial q}~,\label{LegendreD..02} \\
\dfrac{\partial }{\partial \tau}(\Delta f_{{\rm D}m(l)})
&=&\dfrac{q k}{(2l+1)\varepsilon_{{\rm D}m} }\left( 
l \Delta f_{{\rm D}m({l-1})}-(l+1)\Delta f_{{\rm D}m(l+1)} \right)~~~({\rm for}~l \ge 3)~\label{LegendreD..03}.
\end{eqnarray}
By using $\Delta f_{{\rm D}m(0)}$, $\Delta f_{{\rm D}m(1)}$ and $\Delta f_{{\rm D}m(2)}$,   
the perturbed energy density $\delta \rho=\overline{\rho }\delta $, 
perturbed pressure $\delta p=\overline p \pi_{\rm L}$, energy flux $\theta$ 
and shear stress $\sigma$ 
can be written as
\begin{eqnarray} 
\overline{\rho}_{{\rm D}m}\delta_{{\rm D}m}&\equiv &\dfrac{1}{a^{4}}
\displaystyle\int dq\,4\pi q^{2}\varepsilon_{{\rm D}m }\Delta f_{{\rm D}m(0)}~,
\label{vel00}\\
\overline p_{{\rm D}m} \pi_{{\rm LD}m}&=&\dfrac{1}{3a^{4}}
\displaystyle\int dq\,4\pi q^{2}\dfrac{q^{2}}{\varepsilon_{{\rm D}m }}\Delta f_{{\rm D}m(0)}~,\\
\left( \overline{\rho}_{{\rm D}m}+\overline{p}_{{\rm D}m}\right)\theta_{{\rm D}m}
&=&\dfrac{k}{a^{4}}
\displaystyle\int dq\,4\pi q^{2}~q\Delta f_{{\rm D}m(1)}~,\\
\left( \overline{\rho}_{{\rm D}m}+\overline{p}_{{\rm D}m}\right)\sigma_{{\rm D}m}
&=&\dfrac{1}{a^{4}}
\displaystyle\int dq\,4\pi q^{2}\dfrac{q^{2}}{\varepsilon_{{\rm D}m }}\Delta f_{{\rm D}m(2)}~,
\label{vel22}
\end{eqnarray}
where $\overline{\rho}_{{\rm D}m}$ and $\overline{p}_{{\rm D}m}$ are 
the mean energy density and pressure of the $m$-th daughter particles, 
respectively. 
Here, $\pi_{{\rm LD}m}$, 
$\theta_{{\rm LD}m}$ and $\sigma_{{\rm D}m}$ are 
higher-order velocity-weighted quantities 
since phase space integral in 
these are weighted by the velocity $q/\varepsilon$ or velocity squired $(q/\varepsilon)^2$ compared
with the density perturbation $\delta_{{\rm D}m}$.

 Equations (\ref{LegendreD..01})-(\ref{LegendreD..03}) 
are the same as those for massive neutrinos except for the last term in
eq.~(\ref{LegendreD..01}) 
(see ref.~\cite{1995ApJ...455....7M}), which is responsible for the creation of density perturbation 
of the daughter particles from that of the mother ones. 
While the structure of the perturbation equations may  not seem to
differ from that of massive neutrinos significantly,  
in fact their solution is far more complicated.
The complication arises from the source terms (the second and third terms in the right hand side of in eq.~\eqref{LegendreD..01}
and the last term in eq.~\eqref{LegendreD..02}). These terms contain delta functions on $\tau$, 
which clearly {require} a specialized treatment in numerical computation.

In order to solve the perturbation equations and obtain the CMB angular power
spectrum $C_l$ 
as well as the matter power spectrum $P(k)$ 
we modified the publicly available 
{\tt CAMB} code \cite{2000ApJ...538..473L}. 
{In particular, we need to calculate the evolution
of the phase space distribution of the daughter particles $\Delta f_{{\rm D}m(l)}$
at discrete values of $q$.
In the following, we describe how we have chosen the discrete samples of $q$.
}

{
Roughly speaking, {we need to choose} 
the range of $q$ where the mean number 
density of the daughter particles par $q$, 
$\mathcal F(q) \equiv q^2 \bar f(q,\tau_0)$, dominantly contributes 
to the integrals in Eqs.~\eqref{vel00}-\eqref{vel22}.
Let us denote the scale factor at $t=\min(t_0,\Gamma^{-1})$ as $a_{\rm D}$.
Then the location of the maximum of $\mathcal F(q)$
is approximately given by  $q=a_{\rm D}p_{\rm Dmax}$.
In our analysis, we sample $q$ in a range $10^{-4} a_{\rm D}p_{\rm Dmax} \le
q \le \min(p_{\rm Dmax}, 5a_{\rm D}p_{\rm Dmax})$, with a linearly homogeneous spacing in $q$.
Outside this range, $\mathcal F(q)\propto q^{1/2} \exp[-(q/a_{\rm D}p_{\rm Dmax})^{3/2}]$ 
is less than 1/100 of the maximum value, and contributions from such a range of $q$ 
would little affect the CMB and matter power spectra.\footnote{{
We note that undersampling of the phase space of the daughter particles at $a\lesssim 10^{-4}a_{\rm D}$
little affects cosmological observables as the daughter particles are energetically irrelevant
and hardly affect the metric perturbations at this epoch.}}
For a range of the DDM parameters as $0.1\leq m_{\rm D1}\slash m_{\rm M} \leq 1$ and 
0.01~Gyr $\leq \Gamma^{-1} \leq $ 1000~Gyr, which is of our primary interest in this paper, 
we have found that it is sufficient to take the number of sampled $q$ and the maximum multipole $l$ of 
$\Delta f_{{\rm D}m(l)}$ to be 1000 and 45.
We confirm that if we change the number of $q$ to be 1500 and 2000 and/or the maximum $l$ to be 60 and 100, 
the results we will present in Section~\ref{sec:signatures} would differ by no more than 0.8\%.
}

The initial conditions of $\Delta f_{{\rm D}m(l)}(q,\tau)$ 
are set as follows.
When $\tau<\tau_{q}$, both daughter particles 
which have comoving momentum $q$ have never been generated. 
Thus 
\begin{eqnarray} 
\overline{f}_{\rm D}(q,\tau)=\Delta f_{\rm D}(q,\boldsymbol n,\tau)=0~~~({\rm for}~\tau<\tau_q).
\end{eqnarray}
In eqs.~(\ref{LegendreD..01}) and (\ref{LegendreD..02}), 
the source terms $({1}\slash{6})\dot{h}_{\rm L}q
({\partial \overline{f}_{\rm D}}\slash{\partial q})$, 
$({a\Gamma \overline N_{\rm M}}\slash{4\pi q^3 \mathcal H}) \delta_{\rm M} \delta (\tau-\tau_q)$ 
and $\left(({1}\slash{15})\dot{h}_{\rm L}+({2}\slash{5})\dot{\eta}_{\rm T} \right)
q{\partial \overline{f}_{\rm D}}\slash{\partial q}$ contain 
a delta function $\delta (\tau - \tau_{q})$, 
which makes $\Delta f_{{\rm D}m(0)}$ and $\Delta f_{{\rm D}m(2)}$ 
arise like a step function at $\tau=\tau_{q}$. 
In order to treat these terms, we obtain 
the initial values by integrating eqs.~(\ref{LegendreD..01}) and (\ref{LegendreD..02})
with $\tau$ in a infinitesimal interval around $\tau=\tau_{q}$. 
The initial values of $\Delta f_{{\rm D}m(0)}$ and $\Delta f_{{\rm D}m(2)}$ 
are provided in eqs.~(\ref{inil0}) and (\ref{inil2}) 
in Appendix \ref{app:analytic}. 
For $\tau>\tau_q$, $q{\partial \bar f_{\rm D}}\slash{\partial q}$ becomes a smooth function 
of $\tau$ and the time evolutions of the perturbed distribution functions of the daughter particles
can be calculated in the same way as massive neutrinos in the standard cosmology.

\section{Time evolutions of perturbations} \label{sec:evol}

{
In this section, we discuss time evolution{s} of perturbed quantities.
Hereafter we continuously assume that $m_{\rm D2}$ is zero, while the mass of the first 
daughter particles $m_{\rm D1}$ can vary in $[0, m_{\rm M}]$. 
For later convenience, we respectively denote the scale factor at the decay time and the horizon crossing as $a_{\rm d}$
and $a_{\rm hc}$, i.e.,  $t(a_{\rm d})\equiv\Gamma^{-1}$ 
and $\tau(a_{\rm hc})\equiv\pi k^{-1}$. 
}

\subsection{A case of non-relativistic decay}\label{non-relativistic}

{
In this subsection, we consider a case with $m_{\rm D1}\simeq m_{\rm M}$, 
which we refer to as non-relativistic decay. As a representative value, 
we here adopt $m_{\rm D1}=0.999\,m_{\rm M}$, 
with which the massive daughter particles are produced with a velocity kick
$v_{\rm kick}\simeq1-m_{\rm D1}/m_{\rm M}=0.001$. 
We expect time evolutions of perturbations depend on whether the scale of perturbations
is inside or outside the horizon, as well as whether the DDM has decayed or not.
Therefore in what follows we separately investigate cases with $a_{\rm d}<a_{\rm hc}$ 
and $a_{\rm d}>a_{\rm hc}$. 
}

\subsubsection{Perturbations crossing the horizon after the decay time}
\begin{figure}[thb]
\centering
\includegraphics[width=160mm]{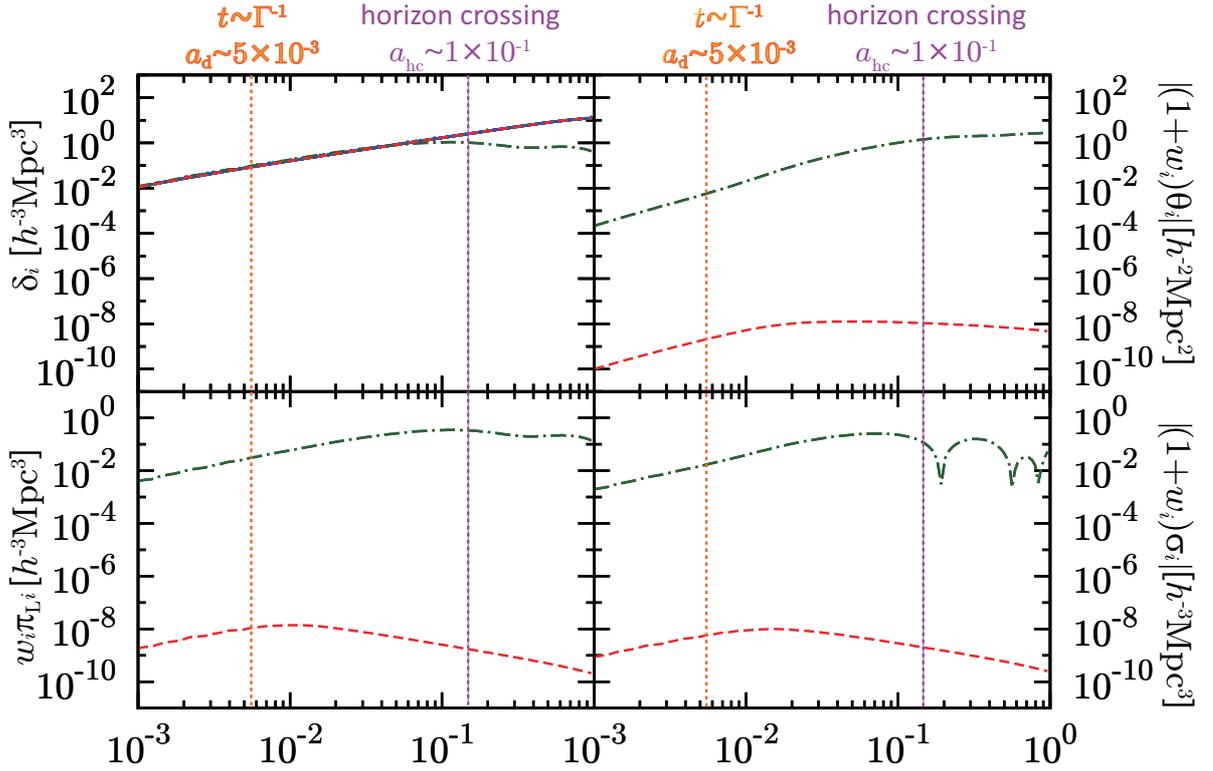}
\caption{
Time evolutions of perturbations in the case of non-relativistic decay.
Here we show perturbations whose scale $k=8\times 10^{-4}~h $Mpc$^{-1}$
crosses the horizon after the decay time $\Gamma^{-1}=6~$Myr.
Shown are the density perturbations (top left), energy fluxes (top right), 
pressure perturbations (bottom left) and anisotropic stresses (bottom right) 
of the mother (blue {solid} line), massive daughter (red {dashed} line) and 
massless daughter (green {dot-dashed} line) particles. 
{In the upper left panel, the line of 
the density perturbation of the massive daughter overlaps 
that of the mother particle.}
Two dotted vertical lines indicate the horizon crossing and the decay time.
$w_{i}=\overline{p}_{i}\slash \overline{\rho}_{i}$ is the equation of state of the $i$-th component.
{According to our gauge choice, 
the dipole and quadrupole moment of the mother particle are zero. 
Thus the energy flux and anisotropic stress of mother particle are zero.
In addition, because of our assumption that 
the momentum of mother particle is negligible compared with its mass, 
the pressure perturbation of mother particle becomes zero. 
Therefore these quantities of mother particle{, such as
$\theta_{\rm M},~\pi_{\rm M}$ and $\sigma_{\rm M}$, } 
are {zero and} not shown {in upper right and 
bottom panels.}}
}
\label{mD1.mM..0.998.T.0.08.superhorizon}
\end{figure}

{
Let us first consider perturbations {which} cross the horizon
after the decay of DDM, i.e. $a_{\rm d}<a_{\rm hc}$. Here we adopt $\Gamma^{-1}=6~$Myr and 
consider evolutions of perturbations at $k=8\times10^{-4}~h$Mpc$^{-1}$. This setup corresponds to 
$a_{\rm d}\simeq 5\times 10^{-3}$ and $a_{\rm hc}\simeq 0.1$.
In figure \ref{mD1.mM..0.998.T.0.08.superhorizon}, we plot time evolutions of perturbation quantities of 
the mother and daughter particles, including density perturbations $\delta_i$, 
pressure ones $\pi_{{\rm L}i}$, energy fluxes $\theta_i$ and shear stresses $\sigma_i$.
Note that regarding perturbed quantities of the mother particles,
only the density perturbation $\delta_{\rm M}$ is nonzero due to
the vanishing momentum distribution of the mother particles and our choice of gauge.
}

Before the decay of DDM $a<a_{\rm d}$, the density perturbations of the daughter particles 
$\delta_{{\rm D}1}$ and $\delta_{{\rm D}2}$ grow in proportion to $\delta_{\rm M}$. 
{
It is because monopole moments $\Delta f_{{\rm D}m(0)}$ are sourced by the density perturbation 
of the mother particles $\delta_{\rm M}$ and the metric perturbation $h_{\rm L}$, 
which also grow during the matter-domination epoch.}
In particular, sufficiently prior to the decay time in the matter-dominated Universe, 
$\delta_{\rm D1}$ and $\delta_{\rm D2}$ are related to $\delta_{\rm M}$ as 
\begin{eqnarray} 
\delta_{\rm D1}&=&\delta_{\rm M}~, \label{massive.daughter.evolution.in.MD}\\
\delta_{\rm D2}&=&\dfrac{23}{21}\delta_{\rm M}~, \label{massless.daughter.evolution.in.MD}
\end{eqnarray}
on superhorizon scales, 
which can be derived analytically as in Appendix \ref{app:analytic}.
{Our numerical result is consistent with analytic one as seen 
in the upper left panel of figure \ref{mD1.mM..0.998.T.0.08.superhorizon}.}

Since the massive daughter particles are non-relativistic, their pressure perturbation{s} 
{are} suppressed as $\delta p_{\rm D1}\simeq \mathcal O(v_{\rm kick}^2) \delta \rho_{\rm D1}$, while
that of the massless daughter one is large as $\delta p_{\rm D2}=\delta\rho_{\rm D2}/3$.
As can be seen in eq.~\eqref{LegendreD..0101} dipole moments $\Delta f_{{\rm D}m(1)}$
are sourced only by monopole and quadrupole moments via free-streaming.
At superhorizon scales $k<\pi \tau^{-1}$, therefore dipole moments $\Delta f_{\rm Dm(1)}$ and 
hence the energy fluxes $\theta_{\rm Dm}$ are little generated. 
On the other hand, as quadrupole moments  {$\Delta f_{\rm Dm(2)}$
are directly generated by metric perturbations (see eq.~\eqref{LegendreD..0101}), 
$\sigma_{\rm Dm}$ can be large.
In addition, for the same reason as the pressure perturbation $\delta p_{\rm D1}$, 
velocity-weighted quantities of the massive daughter particles, including
$\theta_{\rm D1}$ and $\sigma_{\rm D1}$, are further suppressed.
}

After the decay of DDM but still before the horizon crossing $a_{\rm d}<a<a_{\rm hc}$, 
daughter particles are no longer 
sourced by the mother particles 
and the last term of eq.~(\ref{LegendreD..01}) becomes
negligible. 
In this epoch, the Boltzmann equations for the daughter particles 
become the same as that for collisionless free-streaming 
particles, e.g. massive neutrinos (see ref.~\cite{1995ApJ...455....7M}).
For non-relativistic particles, the evolution equation of $\Delta f_{\rm D1}$ should
be effectively reduced to that of CDM. Hence the density fluctuation of the massive daughter particles evolves
in the same way as the mother ones. Due to redshifting of physical momenta,  
higher-order velocity-weighted quantities such as $\pi_{{\rm LD}1}$, $\theta_{{\rm D}1}$ and $\sigma_{{\rm D}1}$ 
decrease afterward. For the massless daughter particles, 
the evolution equation of $\Delta f_{{\rm D}2}$ becomes the same as that of 
massless neutrinos.

{
After the horizon crossing $a>a_{\rm hc}$, 
the daughter particles start free-streaming.
The free-streaming length of the massless daughter particles
equals to the size of horizon and perturbation quantities decay oscillating.}
For massive daughter particles, the density perturbation $\delta_{\rm D1}$ 
continues to grow inside the horizon in the same way as CDM, 
while the velocity-weighted quantities continue to decay due to redshifting of momentum.

\subsubsection{Perturbations crossing the horizon before the decay time}
\begin{figure}
\centering
\includegraphics[width=160mm]{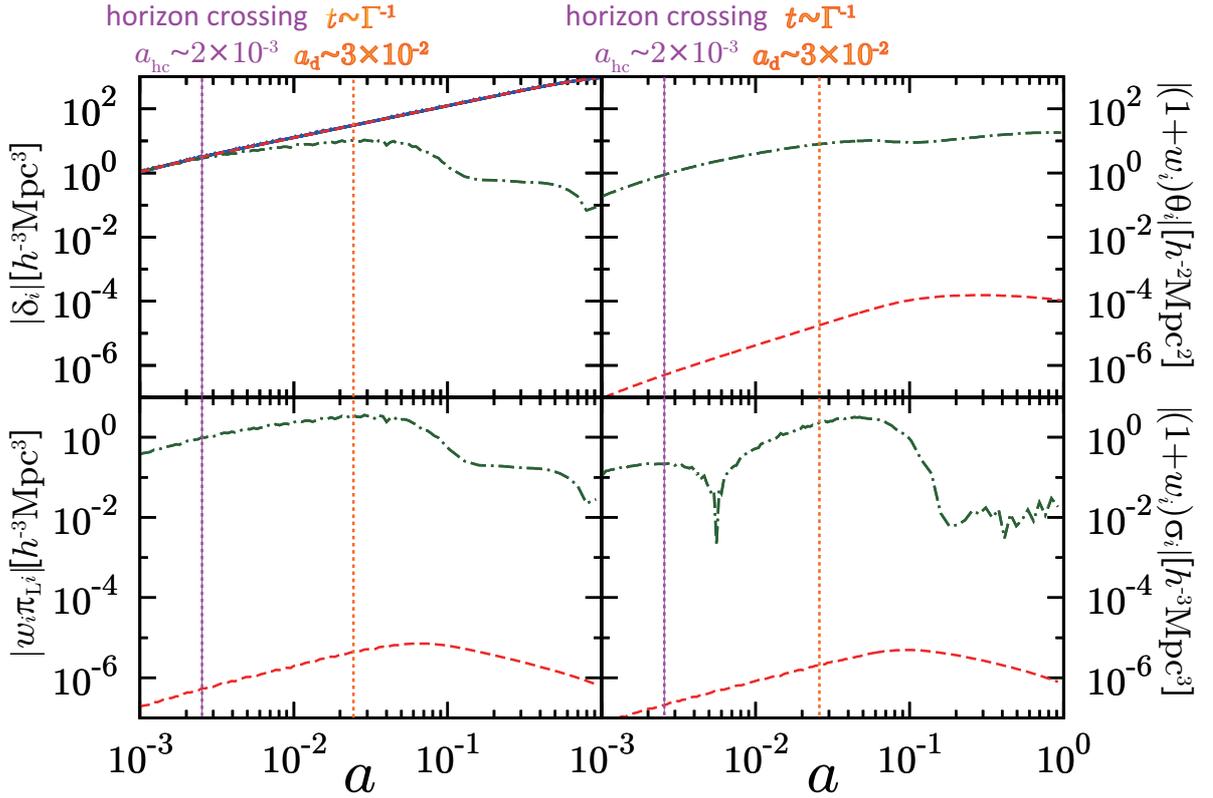}
\caption
{Same as in figure \protect\ref{mD1.mM..0.998.T.0.08.superhorizon} 
but for $\Gamma^{-1}=0.1~$Gyr and $k=8\times 10^{-3}~h$Mpc$^{-1}$. 
{In the upper left panel, the line of 
the density perturbation of the massive daughter overlaps 
that of the mother particle as same as figure \ref{mD1.mM..0.998.T.0.08.superhorizon}.}}
\label{mD1.mM..0.998.T.0.08}
\end{figure}

{ Now we move to a case where DDM decays inside the horizon.  We set the
decay time to $\Gamma^{-1}=0.1~$Gyr, which corresponds to $a_{\rm
d}\simeq 3\times 10^{-2}$, and investigate perturbations on a scale
$k=8\times 10^{-3}~h$Mpc$^{-1}$, which crosses the horizon at $a_{\rm
hc}\simeq 2 \times 10^{-3}$.  Time evolutions of perturbed quantities related to the mother
and daughter particles are plotted in figure \ref{mD1.mM..0.998.T.0.08}}.

{
By comparing {figure \ref{mD1.mM..0.998.T.0.08} with figure \ref{mD1.mM..0.998.T.0.08.superhorizon},} 
we can say that at superhorizon scales when $a<a_{\rm hc}$ and in this case inevitably before the decay, 
behaviors of perturbed quantities of the mother and daughter particles are qualitatively the same 
as in the previous case where the horizon crossing occurs after the decay.
In particular, the analytical solutions of eqs.~\eqref{massive.daughter.evolution.in.MD} and 
\eqref{massless.daughter.evolution.in.MD} again hold at superhorizon scales, which
can be qualitatively confirmed {in the upper left panel of} 
figure \ref{mD1.mM..0.998.T.0.08}.
}

{
On the other hand, evolutions of perturbations inside the horizon
drastically differ from those in
the previous case. After the horizon crossing but still before the decay $a_{\rm hc}<a<a_{\rm d}$, 
the massless daughter particles start to free-stream. However, as we can see in figure \ref{mD1.mM..0.998.T.0.08}, 
the perturbation quantities, such as $\delta_{\rm D2}$,  $\theta_{\rm D2}$ and $\sigma_{\rm D2}$
do not start to oscillate nor decay, contrary to those in figure \ref{mD1.mM..0.998.T.0.08.superhorizon}.
This is because the monopole moment of the massless daughter particles $\Delta f_{{\rm D2}(0)}$ is continuously sourced 
by the density perturbation of the mother particles, and then the dipole and higher multipole moments are also continuously
sourced by $\Delta f_{{\rm D2}(0)}$. Therefore perturbation quantities
of free-streaming relativistic particles keep on growing even inside the horizon. 
Moreover, as for the massive daughter particles, higher-order velocity-weighted quantities including $\pi_{{\rm LD}1}$, $\theta_{{\rm LD}1}$
and $\sigma_{{\rm LD}1}$ also grow. This is because the massive daughter particles with the nonzero velocity $v_{\rm kick}=0.001$ 
keep on being produced, and these particles with relatively large momenta free-stream to make 
higher-order velocity-weighted quantities grow continuously in addition to the density perturbation $\delta \rho_{\rm D1}$.
}

{
Finally at  $a>a_{\rm d}$, when there are few mother particles to decay, 
the source terms in the Boltzmann eqs.~\eqref{LegendreD..01}-\eqref{LegendreD..03}
for the daughter particles become ineffective
and these particles behave as collisionless free-streaming particles. 
Perturbation quantities of the massless daughter particles start to oscillate and decay
due to free-streaming. Comparing figure \ref{mD1.mM..0.998.T.0.08} with figure \ref{mD1.mM..0.998.T.0.08.superhorizon},
we can see perturbation quantities of the massless daughter particles decay more quickly and effects of free-streaming are
more prominent than the previous case. 
This is because most of the daughter particles are produced when $a\simeq a_{\rm d}$
and these particles immediately free-stream over distances larger than the perturbation scale.
On the other hand, since the massive particles are non-relativistic, 
effects of free-streaming is not significant and higher-order velocity-weighted quantities 
decrease mostly due to redshifting of momentum.
}

\subsection{A case of relativistic decay}\label{relativistic}

{
In this subsection, we consider a case with $m_{\rm D1}\ll m_{\rm M}$, 
which we refer to as relativistic decay. 
Here we adopt $m_{\rm D1}={0.1}m_{\rm M}$, which gives
a velocity kick of the massive daughter particles $v_{\rm kick}\simeq
1-2(m_{\rm D1}/m_{\rm M})^2={0.98}$.
In the same way as section  \ref{non-relativistic}, we in the following explore 
time evolutions of perturbations in cases with $a_{\rm d}<a_{\rm hc}$ and 
 $a_{\rm hc}<a_{\rm d}$ separately.}

\subsubsection{Perturbations crossing the horizon after the decay time} \label{sec:rel_late_cross}
\begin{figure}[thb]
\centering
\includegraphics[width=160mm]{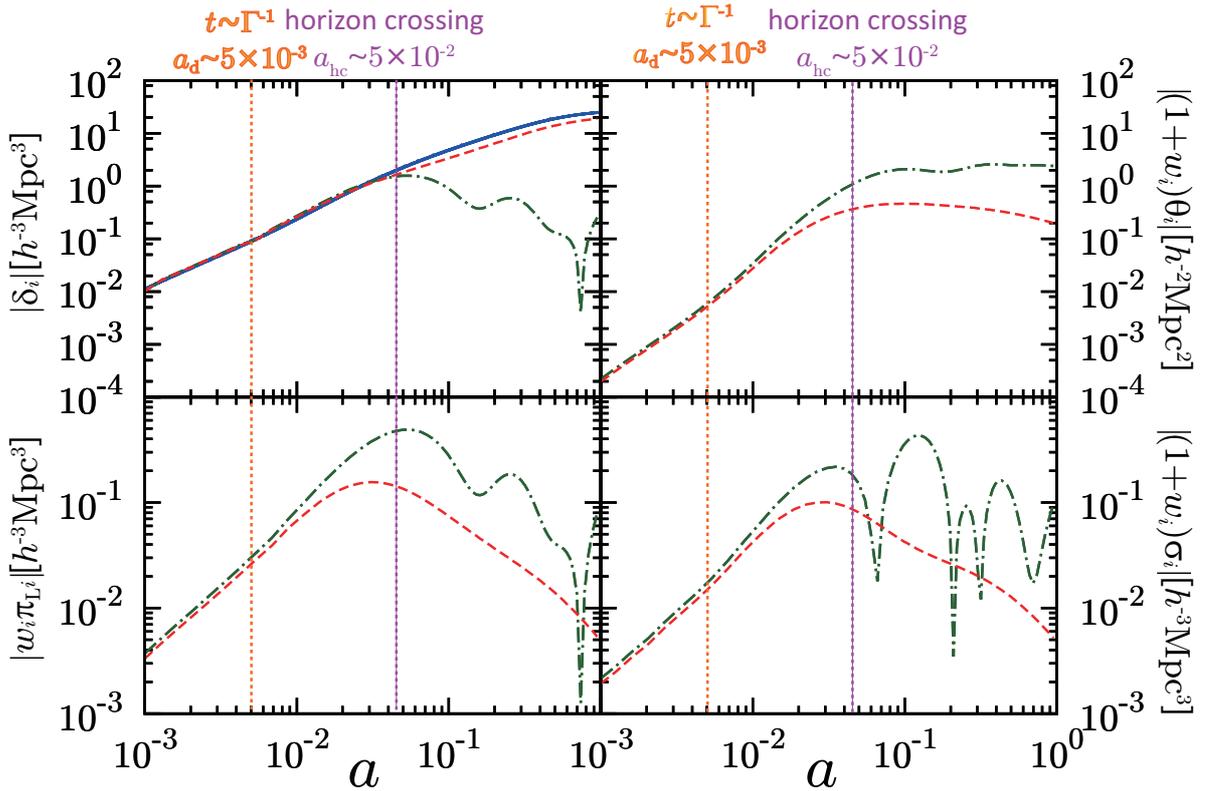}
\caption{Same as in figure \protect\ref{mD1.mM..0.998.T.0.08.superhorizon}
but for $m_{\rm D1}/m_{\rm M}={0.1}$.
}
\label{mD1.mM..0.500.T.0.08.40.0.superHorizon}
\end{figure}

{
Let us see time evolutions of perturbations whose scale crosses the horizon 
after the decay time. We adopt a decay time $\Gamma^{-1}=6~$Myr and
the scale of perturbations is fixed to $k=8\times10^{-4}~h$Mpc$^{-1}$.
These parameters lead to $a_{\rm d}=5\times 10^{-3}$ 
and $a_{\rm hc}=5\times 10^{-2}$.
Figure \ref{mD1.mM..0.500.T.0.08.40.0.superHorizon} shows
time evolutions of perturbation quantities of the mother and daughter particles. 
}

{
First of all, we expect cosmological effects of the two daughter particles are identical as 
long as both of them are sufficiently relativistic. Noting that momenta of particles scale as $a^{-1}$, 
the time when most of the massive daughter particles become non-relativistic can be estimated 
as $a_{\rm nr}= ({v_{\rm kick}}\slash{\sqrt{1-v_{\rm kick}^2}})a_{\rm d}$, which
is roughly ${2.5\times 10^{-2}}$ 
with the parameter values we adopted here. Therefore 
{when} $a<a_{\rm nr}$, perturbation quantities of the two daughter particles are almost the same, 
which can be confirmed in figure \ref{mD1.mM..0.500.T.0.08.40.0.superHorizon}.
}

{
One big difference from the case of non-relativistic decay is that 
the density perturbations  $\delta_i$ grow more rapidly after the decay at superhorizon scales 
as can be seen in {the upper left panel of} 
figure \ref{mD1.mM..0.500.T.0.08.40.0.superHorizon}.
The reason can be understood as follows. Since the decay products are relativistic,  the Universe
is effectively dominated by radiation after the decay.  Therefore, as in
the radiation-dominated epoch, 
density perturbations grow as $a^2$ at superhorizon scales instead of
$a$ in the matter-dominated 
epoch (see eqs.~\eqref{delta.time.evolution} and
\eqref{delta.time.evolution2}). 
As seen in the $\Lambda$CDM model
with low $\Omega_{\rm m}$ (we refer to, {e.g.}, ref.~\cite{2003moco.book.....D}), 
this leads to an enhancement in the matter power spectrum 
at scales which crosses the horizon after the matter-radiation equality.
This issue will be discussed further in section  \ref{sec:pk}.
}

{
{
After horizon crossing, even when $a>a_{\rm nr}$, 
due to the non-vanishing velocity and free streaming of daughter particles, 
at first growth of $\delta_{\rm D1}$ is slower compared with the case of non relativistic decay 
(see also {the upper left panel of} 
figure \ref{mD1.mM..0.998.T.0.08.superhorizon}).
This leads that 
the density perturbation of the mother particles $\delta_{\rm M}$
also grows less, 
for the gravitational potential sourced by
$\delta_{\rm D1}$ decays inside the horizon.
After the massive daughter particles become fully non-relativistic, }
their density perturbation starts to grow as CDM does. 
Other perturbation quantities with higher velocity-weights such as
$\pi_{\rm LD1}$, $\theta_{\rm D1}$ and $\sigma_{\rm D1}$ start 
decreasing almost monotonically without violent oscillations, which is also
seen in cases of non-relativistic decay (see 
{the upper right and bottom panel of} 
figure \ref{mD1.mM..0.998.T.0.08.superhorizon}). 
The massless particles keep on free-streaming and 
their perturbations except for velocity divergence continuously decay}.

\subsubsection{Perturbations crossing the horizon before the decay time}
Figure \ref{mD1.mM..0.500.T.0.08,40.0} shows time evolutions of 
perturbations for a case where the decay occurs inside the horizon.
{
Here we adopt $\Gamma^{-1}=0.1~$Gyr and show perturbations at a scale $k=8\times 10^{-3}~h$Mpc$^{-1}$.
These parameters correspond to $a_{\rm d}\simeq3\times 10^{-2}$ and $a_{\rm hc}=2\times 10^{-3}$.
In this case, most of the massive daughter particles become non-relativistic at around 
$a_{\rm nr}={({v_{\rm kick}}\slash{\sqrt{1-v^2_{\rm kick}}})a_{\rm d}\simeq 0.15}$.}

\begin{figure}[thb]
\centering
\includegraphics[width=160mm]{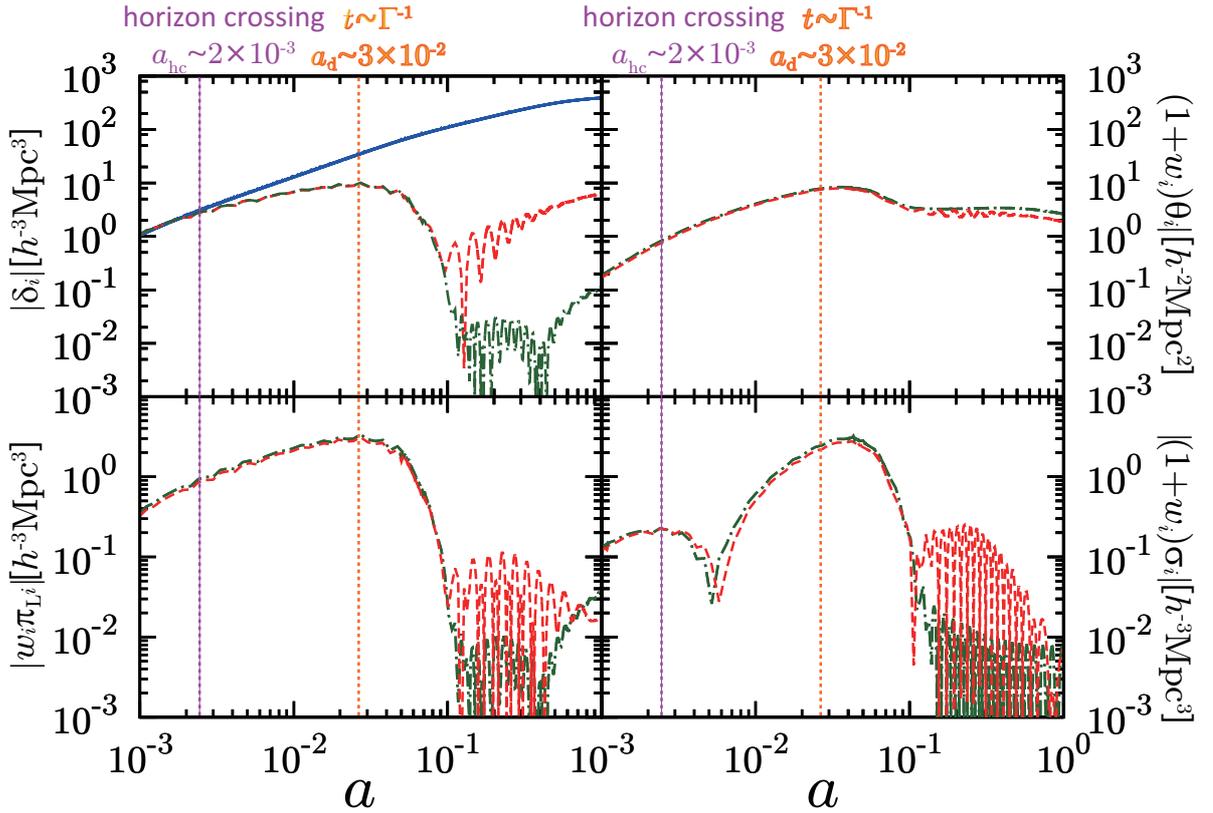}
\caption{Same as in figure \protect\ref{mD1.mM..0.998.T.0.08}, but
for $m_{\rm D1}/m_{\rm M}={0.1}$. 
{Some jaggy features on plots are caused by the numerical error 
on the calculation.}}
\label{mD1.mM..0.500.T.0.08,40.0}
\end{figure}

{
At $a<a_{\rm nr}$, as expected, the perturbation evolutions of the two daughter particles are again
almost the same, which can be seen in  figure \ref{mD1.mM..0.500.T.0.08,40.0}.
As is discussed in the case of non-relativistic decay in section  \ref{non-relativistic}, in this case
perturbation quantities continuously grow inside the horizon until the mother particles completely decay, 
and then effects of free-streaming become significant. This can be confirmed in figure \ref{mD1.mM..0.500.T.0.08,40.0}.
}

{
As the massive daughter particles become non-relativistic at $a>a_{\rm nr}$
their density perturbation $\delta_{\rm D1}$ starts to grow and other velocity-weighted
perturbations $\pi_{\rm LD1}$ etc. start to decrease with less oscillation than before.
However, the decay of $\delta_{\rm D1}$ is so significant, 
{the density perturbation of total matter becomes much smaller than
the density perturbations of the mother particles $\delta_{\rm M}$.
This decay of the density perturbation leads to a significant suppression 
in the matter power at small scales as will be shown in section \ref{sec:pk}.}
On the other hand, the massless daughter particles continuously free-stream.
}

\section{Signatures in cosmological observables}\label{sec:signatures}

\subsection{Effects on $C_{l}$}

\begin{figure}[thb]
\centering
\includegraphics[width=105mm]{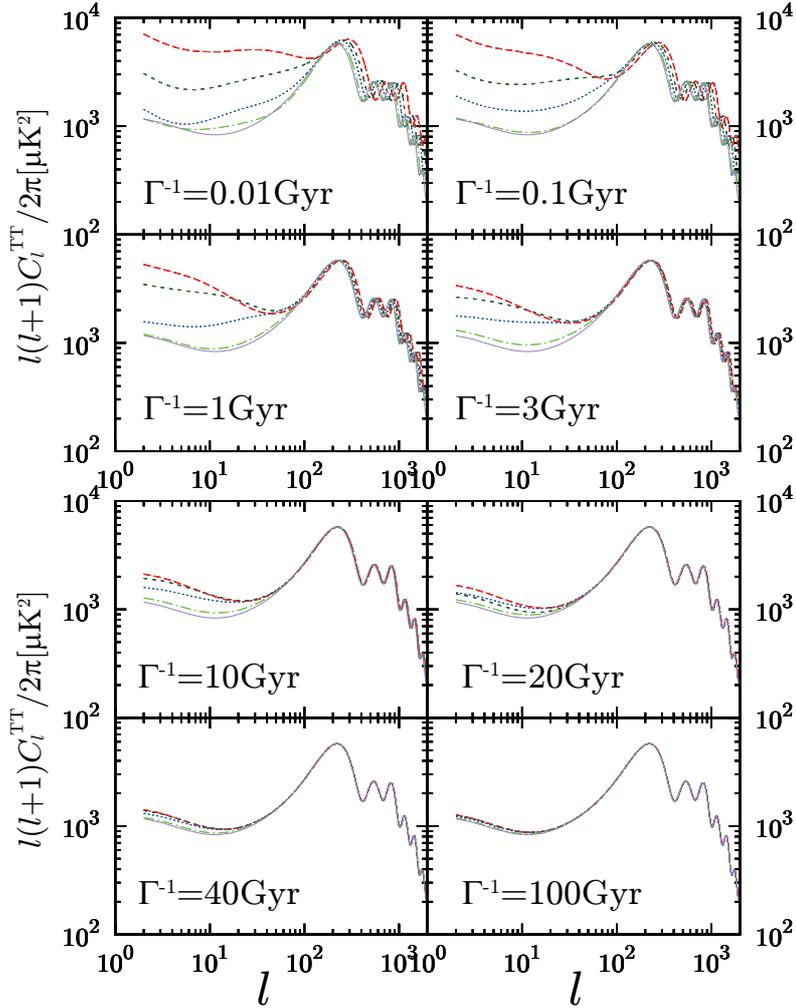}
\caption{Effects of DDM on 
CMB power spectrum of temperature fluctuation $C_{l}^{\rm TT}$. 
Line color distinguishes the mass ratio 
$m_{\rm D1}/m_{\rm M}$. 
Red ({long-dashed}), dark-green ({short-dashed}), 
blue ({dotted}),
green ({dot-dashed}) lines represent 
$m_{\rm D1}\slash m_{\rm M}=0.3,~0.5,~0.7,$ and $0.9$, respectively. 
Purple (solid) lines correspond to $C_{l}^{\rm TT}$ in the $\Lambda $CDM model.
{Each panel shows $C_l^{\rm TT}$ with a distinct decay time $\Gamma^{-1}$, which
varies from $0.01$ (top left) to $100$ Gyr (bottom right)
{as indicated in each panel.}
}
}
\label{cltt}
\end{figure}

In this subsection, we consider the CMB power spectrum
$C_{l}$
{
in the DDM models with $\Gamma^{-1}>t_*$.
In figure \ref{cltt}, we plot the CMB temperature power spectrum} $C_{l}^{\rm TT}$ with 
various decay times $0.01<\Gamma^{-1}\,\rm{[Gyr]}<100$ and mass ratios {$0.3<m_{\rm D1}/m_{\rm M}<0.9$.}
From the figure, we can see effects of DDM
on $C_l^{\rm TT}$ are twofold. First one is the shift
of the positions of the acoustic peaks caused by a change in the background expansion.
{The other} is the integrated Sachs-Wolfe effect induced 
by the decay of the gravitational potential.

Let us {first} investigate the effects on positions of the acoustic peaks.
When $\Gamma^{-1}$ is shorter than the age of the Universe at present $t_{\rm age}\simeq 14~$Gyr 
\cite{2013arXiv1303.5076P} and DDM decays into the relativistic daughter particles, 
energy density in the Universe and hence the expansion rate stay below those in the $\Lambda$CDM model.
This makes angular diameter distances larger and hence the angular size of the sound horizon smaller 
than in the case of the $\Lambda$CDM model. Therefore positions of the acoustic peaks shift toward 
higher $l$.

Second,
let us move to the effects that arise at large angular scales 
larger than the sound horizon at the
recombination epoch.
At these scales, CMB temperature anisotropy can be given as
\cite{1995ApJ...455....7M,1996ApJ...469..437S}
\begin{equation} 
\dfrac{\delta T}{T}\left(\boldsymbol n \right)
=\dfrac{1}{3}\psi(\boldsymbol{x}_{\ast},\tau_{\ast})
+ \displaystyle\int_{\tau_{\ast}}^{\tau_{\rm now }}
d \tau
\left(
 \dot{\phi}(\boldsymbol{x},\tau )
+\dot{\psi}(\boldsymbol{x},\tau )\right)~
~,\label{cmb1}
\end{equation}
{
where $\boldsymbol x=(\tau_{\rm now}-\tau)\boldsymbol n$,
$\phi (\boldsymbol{x},\tau )$ and $\psi (\boldsymbol{x},\tau )$ 
are {the curvature perturbation} and the gravitational potential{, respectively.}
{They} are given in terms of
the metric perturbations in the synchronous gauge as 
$\phi=\eta_{\rm T}-({\mathcal{H}}\slash{2k^{2}})(\dot{h}_{\rm L}+6\dot{\eta}_{\rm T})$
{and
$\psi={1}\slash{2k^{2}}
\left((\ddot{h}_{\rm L}+6\ddot{\eta}_{\rm T})
+\mathcal{H}(\dot{h}_{\rm L}+6\dot{\eta}_{\rm T})
\right)$
}
\cite{1995ApJ...455....7M}.
The second term in eq.~\eqref{cmb1} shows that 
a time derivative of the curvature perturbation $\dot{\phi}$ 
{and that of gravitational potential $\dot{\psi}$}
generate additional CMB temperature fluctuations on large scales, which is called the integrated
Sachs-Wolfe (ISW) effect. As we have shown in section  \ref{sec:evol}, in the DDM models
density perturbations of daughter particles and hence the gravitational potentials 
can decay at $t>\Gamma^{-1}$, due to free-streaming of the daughter particles.
This leads to a large ISW effect. From figure \ref{cltt}, one can find that
the effect is enhanced as $m_{\rm D1}/m_{\rm M}$ decreases, which
can be easily understood as we have seen that the suppression
of density perturbation $\delta_{\rm D1}$ is more significant in case of the relativistic 
decay than in the non-relativistic decay in section  \ref{sec:evol}. 
On the other hand, one can recognize that the largest $l$ where $C_l^{\rm TT}$ is enhanced 
decreases as $\Gamma^{-1}$ increases. This is because the ISW effect is only effective at
scales larger than the free-streaming scale of the daughter particles
within which the
gravitational potential decays. At smaller scales, photons travel
through a number of peaks and troughs
of the gravitational potential and the net effect of the
potential decay becomes negligible. 
}

\begin{figure}[htb]
\centering
\includegraphics[width=105mm]{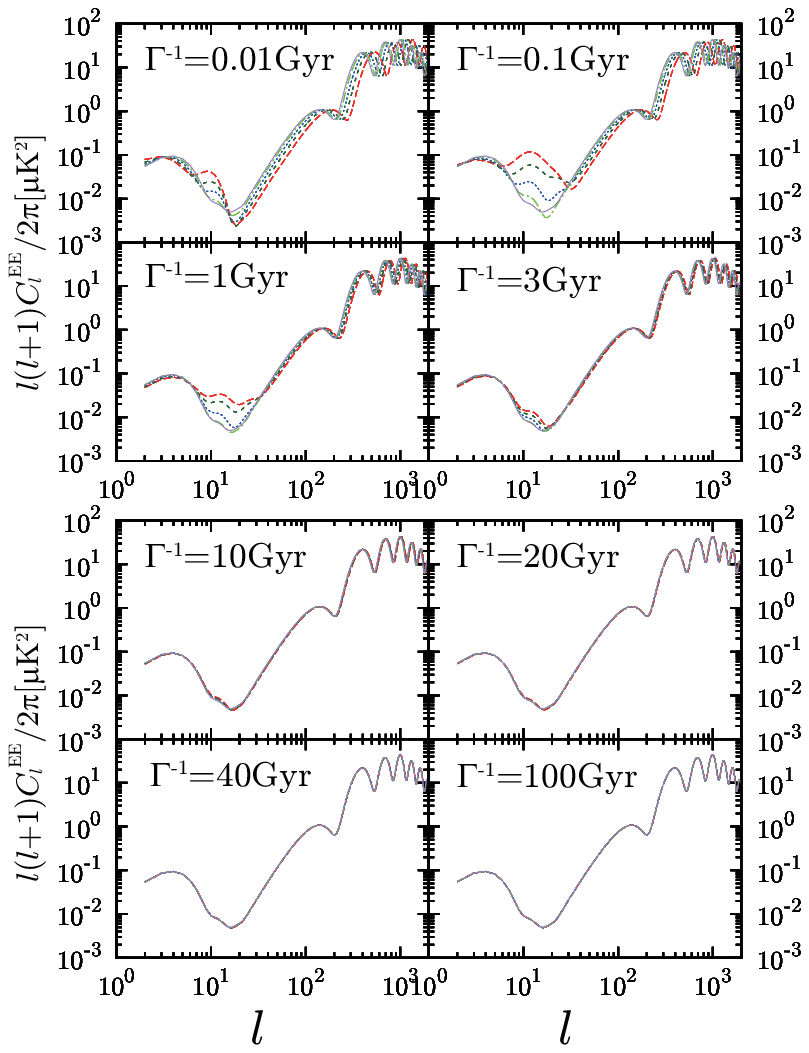}
 \caption{{Same as figure \protect\ref{cltt}
, but for $C_{l}^{\rm EE}$.}}
\label{clEE}
\end{figure}

\begin{figure}[htb]
\centering
\includegraphics[width=105mm]{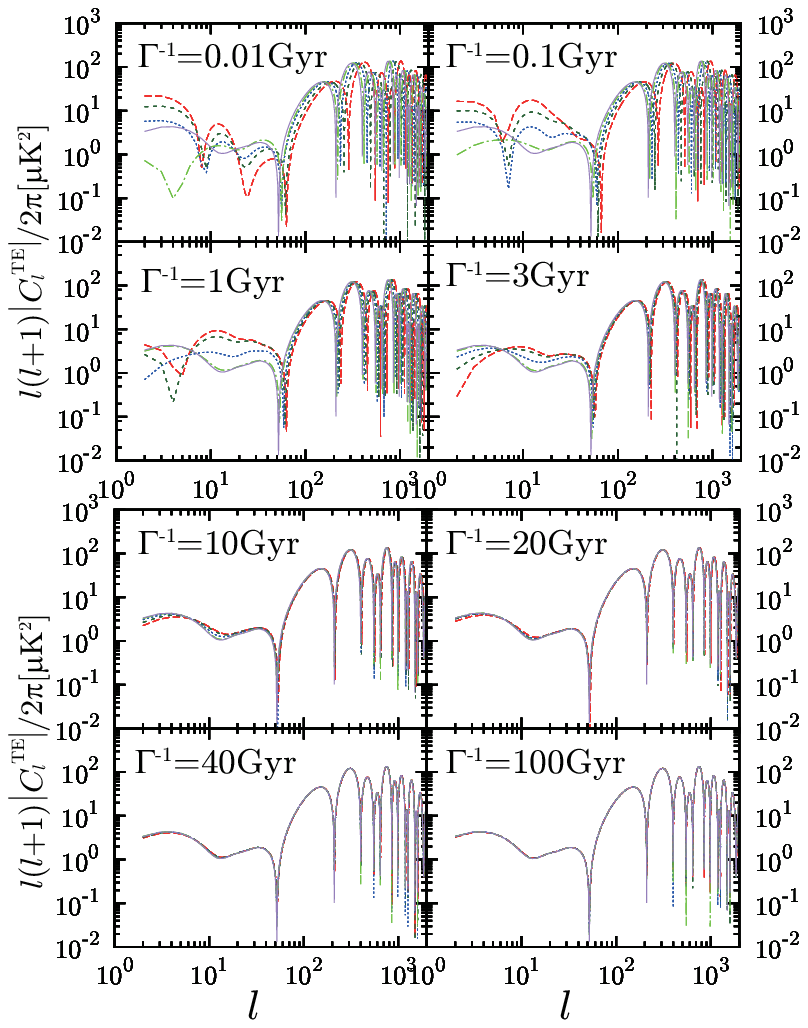}
 \caption{{Same as figure \protect\ref{cltt}
, but for $C_{l}^{\rm TE}$.}}
\label{clTE}
\end{figure}

Finally, let us 
describe signatures in the CMB polarization spectrum 
$C_l^{\rm EE}$ and its cross-correlation with the
temperature anisotropy $C_l^{\rm TE}$. 
In figures \ref{clEE} and \ref{clTE}, 
we plot $C_l^{\rm EE}$ and $C_l^{\rm TE}$, respectively, with the same 
parameter sets as in figure \ref{cltt}.
First, we can see that the DDM models affect $C_l^{\rm EE}$ and $C_l^{\rm TE}$
through the change in the background expansion.
Therefore in the same way as $C_l^{\rm TT}$, 
acoustic peaks and troughs in $C_l^{\rm EE}$ and $C_l^{\rm TE}$
shift toward higher $l$ in the DDM models.
Second, in figures \ref{clEE} and \ref{clTE}
we can also see that there arises an additional E-mode polarization 
{for $\Gamma^{-1} < 3$ Gyr} at $l\sim 10$.
Because {the decay of dark matter} causes an additional ISW effect, 
temperature fluctuations are created at {a} superhorizon scale.
Just after 
horizon crossing, 
the quadrupole moment of temperature fluctuations {is} created 
due to the free-streaming of CMB photons. 
In addition, free-streaming {motion of} 
daughter particles create a nonzero shear stress
and hence induce the anisotropic part of the metric perturbations, which also 
generates the quadrupole moment of the temperature fluctuations.
After the cosmological reionization, 
the E-mode polarization is created 
from the quadrupole moment of 
{the} temperature anisotropy by the Thomson scattering. 
When one considers the epoch well after cosmological recombination,  
the number density of free electrons reaches a maximum 
in the epoch of the cosmological reionization. 
Thus, 
the decay of DDM
generates the additional E-mode polarization 
significantly {if the decay occurs around the reionization epoch }
at $l\sim 10$, which corresponds to the view angle 
of the horizon size 
at the cosmological reionization.

When DDM decays much earlier or much later than the
cosmological reionization epoch, 
the additional quadrupole moment of 
{the} CMB temperature anisotropy at the reionization epoch is small and
the signatures of DDM in $C_l^{\rm EE}$ or $C_l^{\rm TE}$ become insignificant.

\subsection{Effects on $P(k)$}\label{sec:pk}

Let us consider effects of the decay of DDM 
on the matter power spectrum $P(k)$.
As we will see, effects from the decay are more
prominent in $P(k)$ than in $C_l$, 
{so that} we may obtain stronger constraints from observations of $P(k)$
than from those of $C_l$.

\begin{figure}
\centering
\includegraphics[width=105mm]{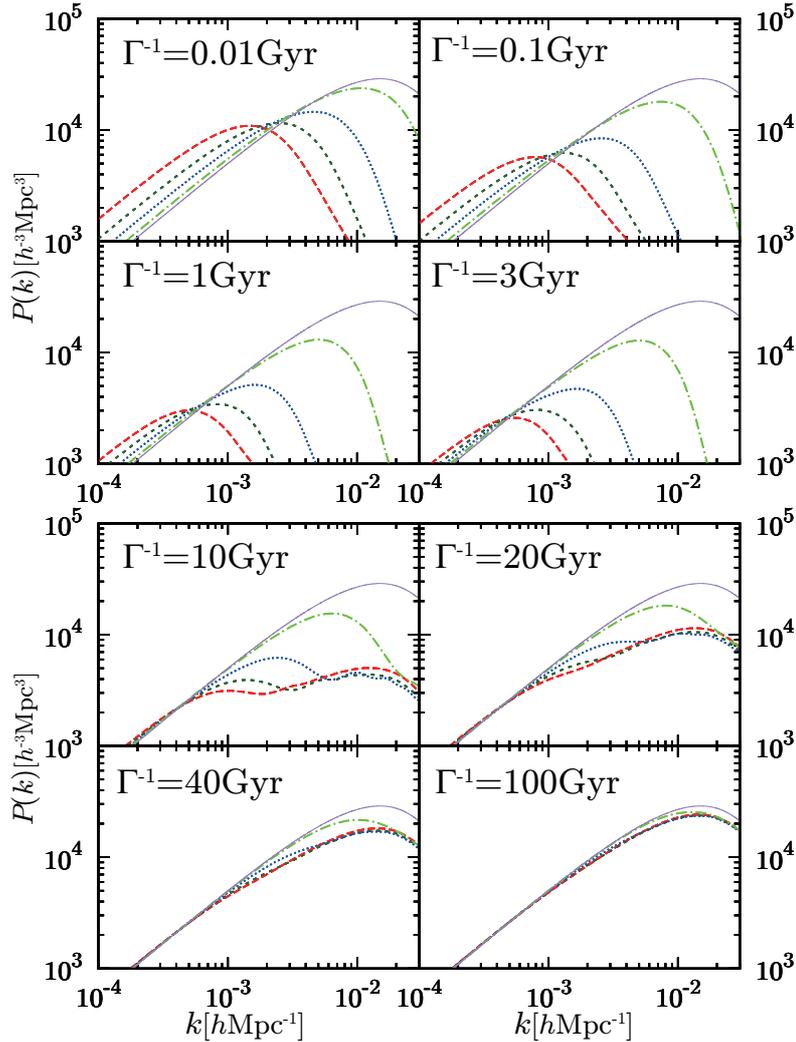}
\caption{$P(k)$ for the DDM model with various parameters 
$(\Gamma^{-1},~m_{\rm D1}\slash m_{\rm M})$. 
Red ({long-dashed}), dark-green ({short-dashed}), 
blue ({dotted}), green ({dot-dashed}) lines represent cases with 
$m_{\rm D1}\slash m_{\rm M}=0.3,~0.5,~0.7,~0.9$, respectively. 
Purple (solid) line corresponds to the $\Lambda $CDM model.
}
\label{Pk}
\end{figure}

In figure \ref{Pk}, we plot the matter power spectra $P(k)$ in the DDM models with
various lifetimes $\Gamma^{-1}$ and mass ratios $m_{\rm D1}/m_{\rm M}$. In the same figure we also 
plot $P(k)$ for the $\Lambda$CDM model as a reference.
From the figure, one can find that there arise two large differences from 
the $\Lambda$CDM model in $P(k)$
. The first one is the suppression at smaller 
scales. This suppression is caused by the free-streaming of the daughter particles, as 
we have discussed in section \ref{sec:evol}. The other one is the enhancement at large 
scales. This enhancement is caused by the growth of density perturbations at superhorizon scales
after the decay of DDM, which we have also discussed in section \ref{sec:rel_late_cross}.
These changes are significant if the decay time is smaller than the age of the Universe and
the mass difference between the mother and daughter particles are large. On the other hand,
if the decay time is larger than the age of the Universe, a significant fraction of the 
mother particles, whose density perturbation grows as that of CDM, still survive 
until today and the deviations in $P(k)$ from the $\Lambda$CDM model become less prominent.

\begin{figure}
\centering
\includegraphics[width=100mm]{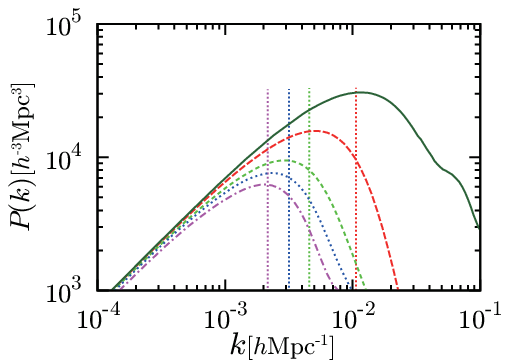}
\caption{Matter power spectrum at present for 
several parameter sets with $\Gamma^{-1}\ll t_{\rm age}$.
On this graph we fixed $m_{\rm D1}/m_{\rm M}$ to $0.7$ . Red ({long-dashed}), 
green ({short-dashed}), blue ({dotted}), 
purple ({dot-dashed}) lines
represent $P(k)$ in the case that the lifetime of DDM 
$\Gamma^{-1}=0.01$, $0.1$, $0.3$, and $1~$Gyr, respectively.
Dotted lines correspond to the free streaming scales 
$k_{\rm FSS}=\pi/\lambda_{\rm FSS}$ 
calculated form eq.~(\protect\ref{FSSeq}) in these parameter sets,  
which correspond to 
2.1, 3.1, 4.7, and 10 [$\times 10^{-3}$ $h$Mpc$^{-1}$], 
respectively. Dark-green line shows $P(k)$ in a case of the $\Lambda $CDM model 
whose dark matter density parameter $\Omega_{\rm c}$ is replaced with 
$m_{\rm D1}\slash m_{\rm M} \times \Omega_{\rm c}$.
Note that these parameter sets 
have been already excluded in our previous work
 \protect\cite{2011JCAP...09..025A}.
}
\label{FSS_Pk}
\end{figure}

To clarify the cause of the changes in $P(k)$, in figure \ref{FSS_Pk}
we plot $P(k)$ for various decay time $\Gamma^{-1}=0.01,~0.1~0.3$ and $1\,$Gyr $\ll t_{\rm age}$ 
with a fixed mass ratio $m_{\rm D1}/m_{\rm M}=0.7$. As a reference, we also plot $P(k)$ {of}
the flat $\Lambda$CDM model in which the CDM density parameter is changed from the fiducial 
value $\Omega_{\rm c}$ to $m_{\rm D1}/m_{\rm M}\times \Omega_{\rm c}$ so that the energy density 
of dark matter in the present Universe should be the same as in the DDM models. 
As we attribute the suppression at small scales to the free-streaming of the daughter particles,
we in addition plot the free-streaming scales of the massive daughter particles $\lambda_{\rm FSS}$, which
can be approximately given as 
\begin{equation}
\lambda_{\rm FSS}=
\displaystyle\int_{\tau_{\rm d}}^{\tau_{0}}d\tau v(\tau)
\sim
\dfrac{v_{\rm kick}}{H_{0}\sqrt{\Omega_{\rm M}}}
\displaystyle\int_{a_{\rm d}}^{1}da
\dfrac{1}{a^{-1/2}\sqrt{q^{2}+m^{2}a^{2}}} 
\sim
\dfrac{3v_{\rm kick}\Gamma^{-1}}{a_{\rm d}}~,
\label{FSSeq}
\end{equation}
where we assumed that the background expansion does not deviate significantly from that 
in the reference $\Lambda$CDM model, which is a good approximation in the case of $m_{\rm D1}/m_{\rm M}=0.7$.
As is expected, we can see that the scales of the suppressions roughly agree with the free-streaming scales of the massive 
daughter particles.
We can also see that the matter power spectra in the DDM models asymptotically become the 
same as that in the reference $\Lambda$CDM model. This shows that the enhancement in $P(k)$ at large scales 
seen in figure \ref{Pk} is explained by the reduction in the energy density of dark matter, which effectively 
changes the matter-radiation equality and growth of the density perturbations at superhorizon scales.

\section{Discussion}\label{sec:discussion}

\begin{figure}
\centering
\includegraphics[width=140mm]{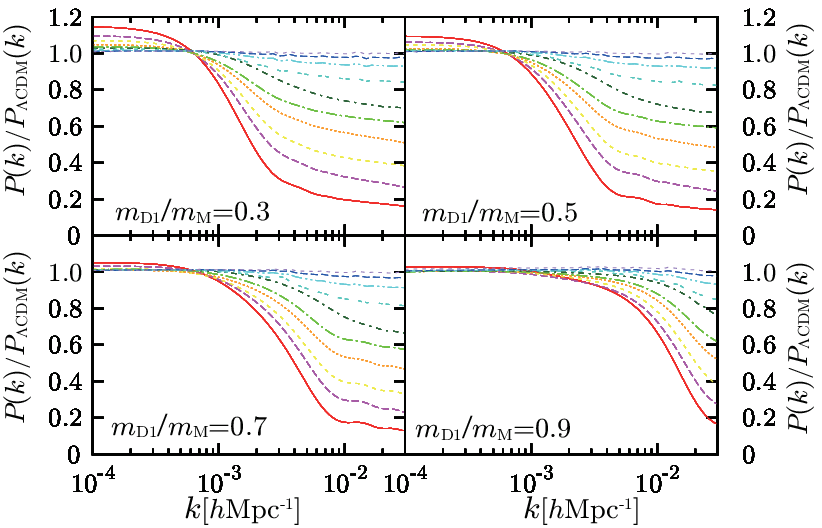}
\caption{The ratio of $P(k)$ to the matter power spectrum in the $\Lambda $CDM model 
$P_{\Lambda{\rm CDM}}$. 
Lines from bottom to top 
represent numerically-obtained values in the case of 
$\Gamma^{-1}=$10, 15, 20, 30, 40, 50, 100, 200, 400, and 800 Gyr,
respectively.
Different panels show $P(k)$ with different mass ratios $m_{\rm D1}/m_{\rm M}$
which are $0.3$ (top left), $0.5$ (top right), $0.7$ (bottom left) and $0.9$ (bottom right).
}
\label{Pk099}
\end{figure}

{In order to illuminate the difference of the power spectrum 
between DDM and $\Lambda$CDM, }
we plot $P(k)$ in the DDM models
normalized by that of the $\Lambda $CDM model, 
$P(k)/P_{\Lambda{\rm CDM}}(k)$ with several parameter sets 
$(\Gamma^{-1}, m_{\rm D1}/m_{\rm M})$
{in figure \ref{Pk099}. }
In the figure we find asymptotic plateaus on small scales,
especially in the plots for small $m_{\rm D1}/m_{\rm M}$ ratios.
{In the small-scale} limit $k\to \infty$, we find
the ratio $r$ to be
\begin{equation} 
r=\dfrac{P(k)}{P_{\Lambda{\rm CDM}}(k)}
\sim \left( \exp\left(-\dfrac{t_{\rm age}}{\Gamma^{-1}} \right)\right)^{2}~.
\end{equation}
Thus, $r$ is responsible for the energy density of 
the surviving mother particles. 
Therefore precise measurements of the matter power spectrum enable us to
distinguish DDM from the warm dark matter (WDM) models, 
because these models predict that the power spectrum 
monotonically decreases to zero 
as 
$k$ increases on smaller scales\footnote{For example, Bode et al. mentions that 
the matter power spectrum in a WDM model 
decreases in proportion to $k^{-10}$ 
on smaller scales \cite{2001ApJ...556...93B}.
}, 
although that of DDM approaches asymptotically to a constant $r$ 
on these scales.

In order to set a constraint on the parameters of the DDM models, 
$\Gamma^{-1}$ and  $m_{\rm D1}/m_{\rm M}$, 
we quantify the effect of DDM on $\sigma_{(R)}$,
which is the fluctuation amplitude at scale $R$ in {units of} $h^{-1}$Mpc.
{Given $P(k)$, }
$\sigma_{(R)}$ can be obtained as

\begin{equation} 
\sigma_{(R)}^{2}= 4\pi
\int_0^\infty
dk\,k^2W^{2}(kR)P(k)~
\label{sigma8},
\end{equation}
where $W(kR)$ is {defined as}
\begin{equation} 
W(kR)\equiv \dfrac{3}{\left( kR \right)^{3}}
\left(\sin(kR)-kR\cos (kR) \right)~,
\end{equation}
{by employing the top hat window function.}

\begin{figure}[h]
\centering
\includegraphics[width=140mm]{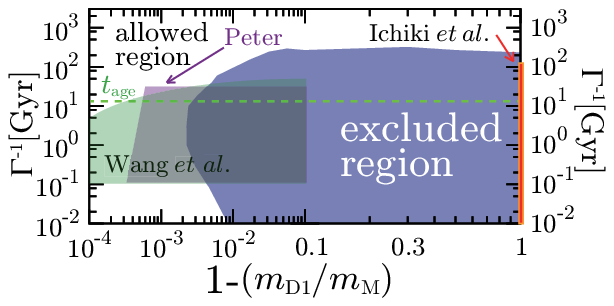}
\caption{Constraint on the lifetime of DDM and the mass ratio of the
 massive daughter particle to the mother particle.  The blue shaded
 region represents the excluded parameter region at 2 $\sigma$
 confidence level from a constraint of $\sigma_{8}$
 \protect\cite{2012MNRAS.425..415S}.  The green ({dashed}) line presents the age
 of the Universe at present $t_{\rm age}=13.8$ Gyr.
The purple{, green} and red shaded regions are the parameter regions 
which Peter \cite{2010PhRvD..81h3511P}, 
{Wang et al. \cite{2013PhRvD..88l3515W}} and
Ichiki et al. \cite{2004PhRvL..93g1302I} have excluded, respectively.} 
\label{FinalConstraint}
\end{figure}

We calculate the fluctuation amplitude at 8 $h^{-1}$Mpc,
$\sigma_{8}$, in the DDM model and compare it to observations \cite{2012ApJ...761...15L,
2012PhRvD..86j3518P,2012MNRAS.425..415S,2012ApJ...756..142V}. 
Recently, $\sigma_{8}$ is reported as
 \cite{2012MNRAS.425..415S}
\begin{equation} 
\sigma_{8}=0.80\pm 0.02~.\label{SDSS3}
\end{equation}
We rule out {the} 
parameter region where $\sigma_{8}^{\rm (th)}$ deviates
from eq.~(\ref{SDSS3}) by more than 2$\sigma$ confidence level 
as shown in figure \ref{FinalConstraint}. 
{
In the same figure, we also depict constraints from 
Peter \cite{2010PhRvD..81h3511P} and Wang et al.\cite{2013PhRvD..88l3515W}, 
which we referred to in introduction.}

{
To understand the constraint in figure \ref{FinalConstraint}, it is convenient to consider cases $\Gamma^{-1}>t_{\rm age}$
and $\Gamma^{-1}<t_{\rm age}$ separately. First, in the case of $\Gamma^{-1}>t_{\rm age}$,
not all of the mother particles have decayed by now. As $\Gamma^{-1}$ increases, more mother particles
survive in the present Universe and the deviation from the $\Lambda$CDM model becomes less.
Therefore, depending on the fractional mass difference $m_{\rm D1}/m_{\rm M}$, small $\Gamma^{-1}$
is excluded. In particular, when the decay is to some extent
relativistic, that is, $m_{\rm D1}/m_{\rm M}\lesssim ${0.9}, 
$\Gamma^{-1}\gtrsim$ {200} Gyr is allowed. We think the reason why the
lower bound on $\Gamma^{-1}$ hardly depends on $m_{\rm D1}/m_{\rm M}$ 
is that the daughter particles have a velocity kick 
which is close to the speed of light. 
This result is consistent with previous works 
\cite{2004PhRvL..93g1302I,2009JCAP...06..005D,
2010PhRvD..81h3511P,2011JCAP...09..025A}. 
On the other hand, when the decay is highly non-relativistic, 
$m_{\rm D1}/m_{\rm M}>${0.998}, the lifetime of DDM is not constrained,
since the evolution of the decay products is indistinguishable from
that of CDM.
}
{
Second, in the case of $\Gamma^{-1}<t_{\rm age}$, all the mother particles
decay into the daughter particles. In this case how $P(k)$ is suppressed 
can be {understood} in terms of the free-streaming length $\lambda_{\rm FSS}$, 
as is discussed in the literature, {e.g.} ref.~\cite{2011JCAP...09..025A}.
Therefore a parameter region with a large mass difference $1-m_{\rm D1}/m_{\rm M}$
is excluded in figure \ref{FinalConstraint}. 
As we have mentioned in section \ref{sec:pk},
given a fixed $m_{\rm D1}/m_{\rm M}$, $\lambda_{\rm FSS}$ decreases
as the mother particles decay earlier and the suppression becomes less significant.
Therefore the constraint on the mass difference 
becomes weaker as the lifetime
$\Gamma^{-1}$ becomes smaller.
}

{Let us remark on constraints from CMB data. As we have shown, DDM models affect the CMB temperature power spectrum effectively in two ways: shifting the angular scale of the acoustic oscillation and enhancing the ISW effect. The constraint from the former effect was derived in Aoyama et al.\cite{2011JCAP...09..025A}, and we have found that it is less strong than one we here derived from $\sigma_8$. On the other hand, we also expect that the latter effect would not be so powerful as $\sigma_8$ due to the cosmic variance.}

We should note that the constraint on $\Gamma^{-1}$ and $m_{\rm D1}/m_{\rm M}$ 
we derived here is overestimated, 
as we fixed other cosmological parameters which may degenerate 
with these two parameters. We will pursue this issue in future works.

\subsection{Implication on the anomaly in estimated $\sigma_{8}$ from Planck}
\begin{figure}[h]
\centering
\includegraphics[width=140mm]{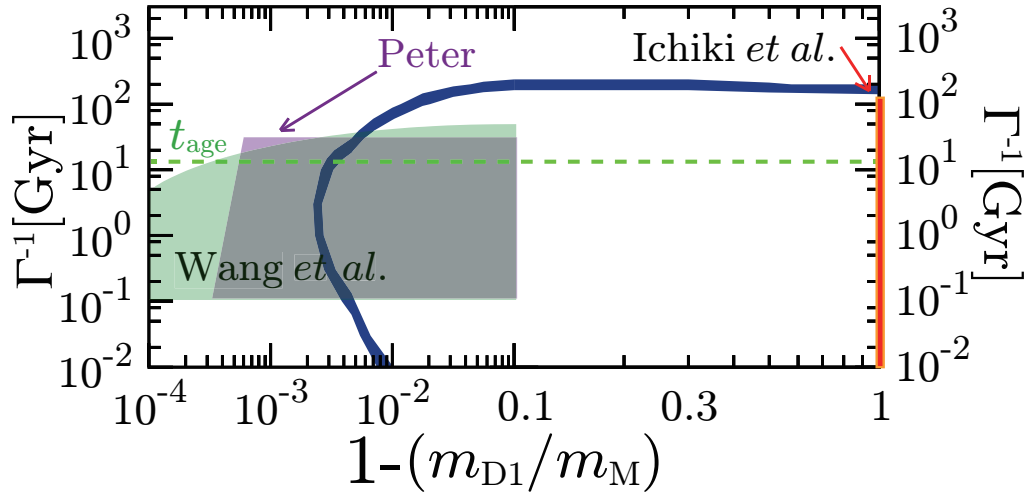}
\caption{The parameter region which can explain the tension in the
 estimated $\sigma_{8}$ 
from the SZ effect cluster number count and from $C_{l}^{\rm TT}$ at 
1$\sigma $ confidence level.
The parameter regions excluded in previous studies are also shown
 in the figure. The purple{, green} and red regions are the same as 
in figure \ref{FinalConstraint}.}\label{planck.constraint}
\end{figure}

Planck collaboration reports that 
the estimated $\sigma_{8}$ from their 
cluster number count through the SZ effect is 
$\sigma_{8}=0.78\pm 0.01$, which is smaller than 
that from the anisotropy of the CMB $\sigma_{8}=0.834\pm 0.027$ 
by more than 2$\sigma$ confidence level \cite{2013arXiv1303.5080P}.
Because the number of clusters reflects 
the matter perturbation in the late-time 
Universe, this discrepancy in the estimated $\sigma_{8}$ may indicate 
some mechanisms which suppress the matter perturbation at small scales 
after cosmological recombination. The DDM model may reconcile the discrepancy, because the decay of DDM with the
lifetime slightly larger than the age of the universe can suppress
the matter power at present keeping the CMB power spectrum almost unchanged.
In figure \ref{planck.constraint}, we plot a  
parameter region 
which can explain the estimated $\sigma_{8}$ obtained from the 
cluster number count
in this DDM model at the 1$\sigma $ confidence level 
with the cosmological parameters obtained by Planck
\cite{2013arXiv1303.5076P}, {i.e.} 
$(\Omega_{\rm b}$, 
$\Omega_{\rm c}$, $h_{\emptyset}$,$\tau_{\rm opt}$, $n_{s}$, $A_{\rm s})$=
(0.04900, 0.2671, 0.6711, 0.0925, 0.9675, $2.215\times 10^{-9}$). 

We see in figure \ref{planck.constraint} that $\Gamma^{-1}\simeq 200$ Gyr is favored 
if $1-m_{\rm D1}/m_{\rm M}\gtrsim 10^{{-1}}$, in which case DDM
decays into two relativistic particles.
On the other hand, 
$\Gamma^{-1}<t_{\rm age}$ is favored if $m_{{\rm D}1}\slash m_{\rm M}\sim 1-10^{-2.5}$, in which case the massive daughter particles are 
non-relativistic when they are produced.
However this parameter  
{region has been} already excluded by Peter
\cite{2010PhRvD..81h3511P} and Wang et al.\cite{2013PhRvD..88l3515W}.
{
Peter is due to observations of the halo mass-concentration and galaxy-cluster mass function.
Wang et al. is due to small scale structures observed by Lyman-$\alpha$ Forest, which should not be destroyed by decaying of dark matter.}

\begin{figure}[h]
\centering
\includegraphics[width=100mm]{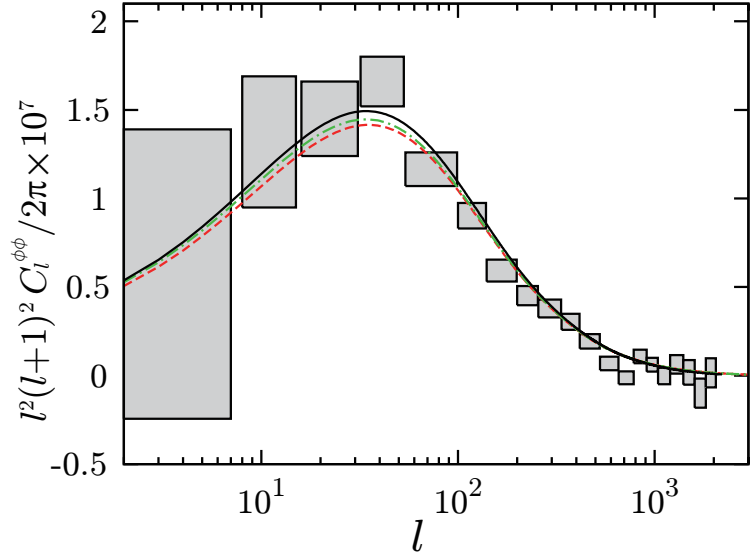}
\caption{{The effect of DDM on the CMB lensing potential power spectrum 
$C_{l}^{\phi \phi}$. The shaded boxes are 
{the data from the Planck experiment \cite{2013arXiv1303.5077P}.}
Red (dashed) and green (dot-dashed) lines 
represent cases with $m_{\rm D1}\slash m_{\rm M}=0.3$ {and} $0.9$, respectively.
In both cases, the lifetime of DDM is set to be $\Gamma^{-1}=200$~Gyr. 
For comparison, the {black} (solid) line represents the case in the 
$\Lambda$CDM model.}
}\label{planck.lensing}
\end{figure}

{
While the DDM model with $\Gamma\gtrsim 200$~Gyr and $1-(m_{\rm D1}/m_{\rm M})\gtrsim0.1$
may be able to solve the discrepancy in $\sigma_8$ estimated from the CMB power spectrum and the SZ cluster counts of Planck, 
one may wonder such models can be constrained by the CMB lens power spectrum $C_l^{\phi\phi}$.
In figure \ref{planck.lensing}, we compare $C_{l}^{\phi \phi}$ in the DDM model 
with the observational result which was reported by the Planck paper \cite{2013arXiv1303.5077P}.
The figure shows that the dark matter decaying suppresses $C_{l}^{\phi \phi}$ 
compared with that in the $\Lambda$CDM model. 
However, it seems that the current data is not constraining enough
for such the parameter region to be excluded.
We defer a more quantitative analysis to future work.
}

\section{Conclusion}\label{sec:conclusion}

In this paper, we considered cosmological consequences of the DDM model where 
the cold mother particles decay into massive and massless particles. In particular, we focused on evolutions of cosmological
perturbations in the model and their signatures in the CMB 
power spectrum $C_{l}$ and the matter power spectrum $P(k)$. 
While similar kinds of models had been studied by various authors, 
we for the first time explored cases with an arbitrary mass ratio between the daughter and mother particles 
$m_{\rm D1}/m_{\rm M}$. For this purpose, we solved the phase space distributions of the 
decay products.
To summarize, the main effect of the decay of DDM is that the free-streaming of the daughter particles 
suppresses structure formation at scales smaller than the free-streaming length.

As for CMB, the DDM model
mainly affects the temperature anisotropy at large angular scales
through the ISW effect. 
{A constraint on $\Gamma^{-1}$ and $m_{\rm D1}\slash m_{\rm M}$ 
from the peak shift of $C_{l}^{\rm TT}$ has been set 
on Aoyama et al.\cite{2011JCAP...09..025A}. 
Their constraint from it is $\Gamma^{-1} >30~$Gyr at $m_{\rm D1}\slash m_{\rm M}\ll 1$}.
However, since measurements of $C_l$ at
large angular sales are fundamentally limited by the cosmic variance, CMB may not be a promising 
probe of the DDM model.
On the other hand, the matter power spectrum  $P(k)$ is 
affected by dark matter decaying significantly even when $\Gamma^{-1}\gg t_{\rm age}$. 
{
Indeed, by using the observational data of $\sigma_{8}$
\cite{2012MNRAS.425..415S}, 
we succeeded in excluding parameter region in the 2D 
plane of the fractional mass difference between the daughter and 
mother particles and the decay time.
If the decay product is relativistic, the decay time $\Gamma^{-1}$
should be longer than 200 Gyr, and if the 
decay time is  shorter than the age of the Universe, 
the fractional mass difference $1-m_{\rm D1}/m_{\rm M}$ 
should be smaller than {$10^{-2.5}$} 
}
The tension between estimated $\sigma_{8}$ from the SZ effect and {the CMB angular power spectrum} 
in the recent Planck data may be explained by the DDM model if $\Gamma^{-1}$ is around {200} Gyr {
and the decay products are relativistic}.

\section*{Acknowledgments}

We thank Masahiro Takada 
for discussions and valuable suggestions. T.S. is supported by the Academy of Finland grant 1263714.
This work is supported in part by scientific research expense for Research Fellow of the Japan Society for the Promotion
of Science from JSPS (24009838)(S.A.) and (23005622) (T.S.)
and Grant-in-Aid for Scientific Research Nos. 22012004
(K.I.), 25287057 (N.S.) of the Ministry of Education, Sports, Science and
Technology (MEXT) of Japan, and also supported by Grant-in-Aid for the Global Center of
Excellence program at Nagoya University "Quest for Fundamental Principles in the Universe:
from Particles to the Solar System and the Cosmos" from the MEXT of Japan. This research
has also been supported in part by World Premier International Research Center Initiative,
MEXT, Japan. 

\appendix

\section{Initial condition for the perturbed distribution functions of the daughter particles}
\label{app:analytic}

{In this appendix,  we first derive the solution of the 
perturbed phase space distribution $\Delta f_{{\rm D}m(l)}$ at $\tau=\tau_q$.
This gives the initial condition needed in solving the perturbation evolution
numerically. As a side product, we then also derive the analytic solution of the 
density perturbations of the daughter particles 
at superhorizon scales well before the decay time.}

{
For later convenience, let us rewrite the solution eq.~\eqref{fD0} for the background 
distribution function $\overline{f}_{\rm D}(q,\tau)$ as }
\begin{equation} 
\overline{f}_{\rm D}(q,\tau)\equiv F_{\rm D}(q){\Theta (ap_{\rm Dmax}-q)}~,
\end{equation}
{
where 
\begin{eqnarray} 
F_{\rm D}(q)
=\dfrac{a_{q}\Gamma \overline{N}_{\rm M}(\tau_q)}
{4\pi q^{3} \mathcal{H}_{q}}~, \label{fDcontinuous}
\end{eqnarray}
which is a smooth function of $q$ and does not depend on $\tau$.
When the decay of the mother particles can be neglected, 
their comoving number density can be approximated as $\overline N_{\rm M}(\tau)=
\overline N_{\rm M\emptyset}$. Thus at $t\ll \Gamma^{-1}$ we can approximate eq.~\eqref{fDcontinuous}
as 
\begin{equation}
F_{\rm D}(q)
\approx\dfrac{a_{q}\Gamma \overline{N}_{\rm M\emptyset}}
{4\pi q^{3} \mathcal{H}_{q}}~.\label{fDapprox}
\end{equation}
}
{
Adopting eq.~\eqref{fDcontinuous},
eq.~\eqref{LegendreD..01} can be rewritten as
\begin{eqnarray}
\frac{\partial \Delta f_{{\rm D}m(0)}}{\partial \tau}=
-\frac{qk}{\epsilon_{{\rm D}m}}\Delta f_{{\rm D}m(1)}
+\frac{\dot h_{\rm L}}{6}q\frac{d F_{\rm D}}{dq}\Theta(\tau-\tau_q)
+\left[-
\frac{\dot h_{\rm L}}{6\mathcal H}+\delta_{\rm M}
\right]F_{\rm D}\delta(\tau-\tau_q)~.\label{l0}
\end{eqnarray}
By integrating eq.\eqref{l0} in an infinitesimal interval around $\tau=\tau_q$, we obtain
\begin{equation}
\Delta f_{{\rm D}m(0)}(q,\tau_q)=
\left[-\frac{\dot h_{\rm L}(\tau_q)}{6\mathcal H_q}+\delta_{\rm M}(\tau_q)
\right]F_{\rm D}(q)~.\label{inil0}
\end{equation}
In the same way, eq.~\eqref{LegendreD..02} can be  rewritten as 
\begin{eqnarray}
\frac{\partial \Delta f_{{\rm D}m(2)}}{\partial \tau}&=&
\frac{qk}{(2l+1)\varepsilon_{{\rm D}m}}\left[l\Delta f_{{\rm D}m(1)}-
(l+1)\Delta f_{{\rm D}m(3)}\right] \notag\\
&&\quad-\left[\frac{\dot h_{\rm L}}{15}+\frac{2\dot\eta_{\rm T}}{5}
\right]q\frac{d F_{\rm D}}{dq}\Theta(\tau-\tau_q)
+\frac1{\mathcal H}
\left[\frac{\dot h_{\rm L}}{15}+\frac{2\dot\eta_{\rm T}}{5}
\right]F_{\rm D}
\delta(\tau-\tau_q)~, \label{l2}
\end{eqnarray}
which leads to
\begin{equation}
\Delta f_{{\rm D}m(2)}(q,\tau_q)=\frac1{\mathcal H_q}
\left[\frac{\dot h_{\rm L}(\tau_q)}{15}+\frac{2\dot\eta_{\rm T}(\tau_q)}{5}
\right]F_{\rm D}(q)~.\label{inil2}
\end{equation}
On the other hand, for $l\ne 0,~2$, as the right hand side of Eqs .\eqref{LegendreD..01} 
and \eqref{LegendreD..03} are smooth around $\tau=\tau_q$, we obtain
\begin{equation}
\Delta f_{{\rm D}m(l)}(q,\tau_q)=0 \quad({\rm for}~l\ne 0,~2).
\end{equation}
}

\subsection{Radiation-dominated era }
{
Now let us consider the superhorizon solution well before the decay time.
In radiation-dominated era, $\delta_{\rm M}$ at superhorizon scales, 
$\mathcal{H}$ and $a$
can be related to $\tau$ as follows 
(see {e.g.} ref.~\cite{1995ApJ...455....7M}).
\begin{eqnarray} 
\delta_{\rm M}&\propto &\tau^{2} ~,\label{delta.time.evolution} \\
\mathcal{H}&= &\tau^{-1}~,\label{conformalHubble} \\
a&\propto &\tau~. \label{atau}
\end{eqnarray}
By substituting eqs.~(\ref{delta.time.evolution}) and \eqref{conformalHubble}
into eq.~(\ref{deltaM.eq.0.5..h})
one can derive a relation 
\begin{equation} 
\dot h_{\rm L}=-2\dot \delta_{\rm M}
=-4\mathcal{H}\delta_{\rm M}~.
\label{deltaM.time-derivative.analitic}
\end{equation}
Substituting eqs.~\eqref{deltaM.time-derivative.analitic} into eq.~\eqref{inil0}, we obtain
\begin{equation} 
\Delta f_{{\rm D}m(0)}(q,\tau_q)=\dfrac{5}{3}F_{\rm D}(q)\delta_{{\rm M}}(\tau_{q})
.~\label{jump2}
\end{equation} 
}
{
Furthermore, by substituting eq.~(\ref{conformalHubble}) into eq.~(\ref{fDapprox}) and $\tau_{q}\propto q$, 
one can find 
\begin{eqnarray} 
F_{\rm D}(q)&\propto& q^{-1}~,\label{fDnormal} \\
q\dfrac{d F_{\rm D}}{d q}&=&-F_{\rm D}~\label{fDdifferential}.
\end{eqnarray}
Then in the limit of $k\tau\to0$, eq.~\eqref{l0} in conjunction with eq.~\eqref{jump2}
gives
\begin{eqnarray} 
\Delta f_{{\rm D}m(0)}(q,\tau)
&=&\left(
\left[ \dfrac{1}{6}h_{\rm L}q\dfrac{\partial F_{\rm D}}{\partial q} 
\right]^{\tau}_{\tau_{q}}
+\dfrac{5}{3}F_{\rm D}(q)\delta_{{\rm M}}(\tau_{q})
\right)\Theta (ap_{\rm Dmax}-q)
\nonumber \\
&=&F_{\rm D}(q)
\left[\dfrac{1}{3}
\delta_{\rm M}(\tau)
+\dfrac{4}{3}\delta_{{\rm M}}(\tau_{q})
\right]\Theta (ap_{\rm Dmax}-q)~.
~\label{6.100}
\end{eqnarray}
Using eqs.~\eqref{delta.time.evolution} and \eqref{atau}, we then obtain
\begin{equation}
\Delta f_{{\rm D}m(0)}(q,\tau)=
\dfrac{1}{3}F_{\rm D}(q) \delta_{\rm M}(\tau)
\left[1+4\left(\frac{a_q}a\right)^2
\right]
\Theta (ap_{\rm Dmax}-q)~.\label{fd.Psi}
\end{equation}
}
Note that this result does not depend on the particle index $m$.
{
By substituting eq.~\eqref{fd.Psi} into eq.~\eqref{vel00}, we can compute the density perturbation
$\delta_{{\rm D}m}$.
When the decay products are relativistic $p_{\rm Dmax}\gg m_{{\rm D}m}$, 
we obtain
\begin{equation} 
\delta_{\rm Dm}(\tau)=\dfrac{17}{15}\delta_{\rm M}(\tau)~. \label{17.15.deltaM}
\end{equation}
Finally, when the decay products are non-relativistic $p_{\rm Dmax}\ll m_{{\rm D}m}$,  
we obtain
\begin{equation} 
\delta_{\rm Dm}(\tau)=\delta_{\rm M}(\tau)~.
\end{equation}
}

 \subsection{Matter-dominated era }
{
In the matter-dominated era well before the decay time, 
$\delta_{\rm M}$ at superhorizon scale,
$\mathcal{H}$ and 
$a$
can be related to $\tau$ as 
\cite{1995ApJ...455....7M}
\begin{eqnarray} 
\delta_{\rm M}(\tau)&\propto &\tau^{2} ~, \label{delta.time.evolution2}\\
\mathcal{H}&= &2\tau^{-1}~, \label{conformalHubble2}\\
a&\propto &\tau^{2}~. \label{atau2}
\end{eqnarray}
Then one can find eq.~\eqref{deltaM.time-derivative.analitic}
should be replaced with
\begin{equation}
\dot h_{\rm L}=-2\mathcal H \delta_{\rm M}, 
\end{equation}
with which eq.~\eqref{jump2} should be replaced with
\begin{equation}
\Delta f_{{\rm D}m(0)}(q,\tau_q)=\dfrac{4}{3}F_{\rm D}(q)\delta_{{\rm M}}(\tau_{q})
.~\label{jump22}
\end{equation}
}
{
By substituting eqs.~(\ref{conformalHubble2}) and \eqref{atau2} 
into eq.~(\ref{fDapprox}), we obtain
\begin{eqnarray} 
F_{\rm D}&\propto& q^{-3/2}~,\label{fDnormal2} \\
q\dfrac{d F_{\rm D}}{d q}&=&-\frac32F_{\rm D}~\label{fDdifferential2}.
\end{eqnarray}
Then from eqs.~\eqref{l0} and \eqref{jump22}, in the limit $k\tau\to0$ we obtain
\begin{equation}
\Delta f_{{\rm D}m(0)}(q,\tau)=
\dfrac{1}{2}F_{\rm D}(q) \delta_{\rm M}(\tau)
\left[1+\frac{5a_q}{3a}\right]
\Theta (ap_{\rm Dmax}-q)~.\label{fd.Psi2}
\end{equation}
In the same way as in radiation-dominated epoch, by substituting eq.~\eqref{fd.Psi2} into
eq.~\eqref{vel00}, we can compute the density perturbation $\delta_{{\rm D}m}$.
When they are relativistic, we obtain {
\begin{equation} 
\delta_{\rm Dm}=\dfrac{23}{21}\delta_{\rm M}~,
\end{equation}
while when the decay products are non-relativistic, we obtain}
\begin{equation} 
\delta_{\rm Dm}=\delta_{\rm M}~.
\end{equation}
}

\bibstyle{JHEP-2}
\bibliography{reference2012a}
\end{document}